\documentclass[fleqn,usenatbib,useAMS]{mnras}

\usepackage{graphicx}	% Including figure files
\usepackage[fleqn]{amsmath}	% Advanced maths commands
\usepackage{amssymb}	% Extra maths symbols
\usepackage{multicol}        % Multi-column entries in tables
\usepackage{multirow}
\usepackage{array}
\usepackage{bm}		% Bold maths symbols, including upright Greek
\usepackage{pdflscape}	% Landscape pages
\usepackage[table]{xcolor}
\usepackage{natbib}
\usepackage{float}
\usepackage[T1]{fontenc}
\usepackage{ae,aecompl}

\usepackage{newtxtext,newtxmath}
\usepackage{comment}
\title[Dynamical parameters of interacting galaxies using CNN]{Determination of the relative inclination and the viewing angle of an interacting pair of galaxies using  convolutional neural networks }

\author[Prakash, Banerjee \& Perepu]{Prem Prakash$^{1}$\thanks{E-mail: \href{mailto:prmprakash163@googlemail.com}{prmprakash163@googlemail.com}}, Arunima Banerjee  $^{1}$\thanks{E-mail: \href{mailto:arunima@iisertirupati.ac.in}{arunima@iisertirupati.ac.in}}, Pavan Kumar Perepu $^{1}$\thanks{E-mail: \href{mailto:pavankumar@iisertirupati.ac.in}{pavankumar@iisertirupati.ac.in}}
\\
$^{1}$Indian Institute of Science Education and Research, IISER Tirupati, India}

\date{}

\pubyear{2020}

\begin{document}
\label{firstpage}
\pagerange{\pageref{firstpage}--\pageref{lastpage}}
\maketitle

\begin{abstract}

Constructing dynamical models for interacting galaxies constrained by their observed structure and kinematics crucially depends on the correct choice of the values of their relative inclination ($i$) and viewing angle ($\theta$) (the angle  between the line of sight and the normal to the plane of their orbital motion). We construct Deep Convolutional Neural Network (DCNN) models to determine the $i$ and $\theta$ of interacting galaxy pairs,  using N-body $+$ Smoothed Particle Hydrodynamics (SPH)  simulation data from the GALMER database for training. GalMer simulates only a discrete set of $i$ values ($0^{\circ}, 45^{\circ}, 75^{\circ} \text{ and } 90^{\circ}$) and almost all possible values of $\theta$ values in the range, $[-90^{\circ}, 90^{\circ}]$. Therefore, we have used classification for $i$ parameter and regression for $\theta$. In order to classify galaxy pairs based on their $i$ values only, we first construct DCNN models for  (i) 2-class ($i$ = 0 $^{\circ}$, 45$^{\circ}$) (ii) 3-class ($i = 0^{\circ},45^{\circ}, 90^{\circ}$) classification, obtaining $F_1$ scores of 99\% and 98\% respectively. Further, for a classification based on both $i$ and $\theta$ values, we develop a DCNN model for a 9-class classification using different possible combinations of $i$ and $\theta$, and the $F_1$ score was 97$\%$. 
To estimate $\theta$ alone, we have used regression, and obtained a mean squared error value of 0.12. Finally, we also tested our DCNN model on real data  from Sloan Digital Sky Survey. Our DCNN models could be extended to determine additional dynamical parameters, currently determined by trial and error method.

\end{abstract}


%%%%%%%%%%%%%%%%%%%%%%%%%%%%%%%%%%%%%%%%%%%%%%%%%%

%%%%%%%%%%%%%%%%% BODY OF PAPER %%%%%%%%%%%%%%%%%%

% The MNRAS class isn't designed to include a table of contents, but for this document one is useful.
% I therefore have to do some kludging to make it work without masses of blank space.
\begingroup
\let\clearpage\relax
%\tableofcontents
\endgroup
\newpage

\section{Introduction}
According to the modern cosmological paradigm, interacting galaxies constitute the building blocks of the hierarchical structure formation of the universe. It also implies formation of massive or giant galaxies due to the merger of dwarf galaxies, which is corroborated by the abundance of early-type galaxies at higher red-shifts, which are marked by high rates of galaxy interaction. Galaxy interactions are further responsible for driving dynamical evolution thus regulating astrophysical phenomena like morphological transformations, starbursts, bulge creation, halo formation etc. and only a small fraction of galaxies evolve in isolation via the pathway of secular evolution (See \cite{Barnes1992} for a review).

One of the pioneering studies in the field of interacting galaxies was done by \cite{Holmberg1937}. \cite{Vorontsov1959} presented a catalog of 355 pairs of interacting galaxies as observed by the Palomar Observatory Sky Survey (POSS). \cite{arp1966} attributed the observed peculiarities and distortions present in these galaxies  to interactions or mergers, studying a total of 338 interacting pairs. \cite{Hibbard2010} studied interacting galaxies at different wavelengths and published a catalog of peculiar galaxies.

Galaxy interaction is a complex dynamical problem, and  therefore not analytically tractable in general. N-body simulation of interacting galaxies was proposed by \cite{Holmberg1941}, who simulated the tidal capture in an interacting pair of galaxies. However, this model did not consider alternative scenarios like repeated encounters for the origin of galaxy mergers. \cite{Zwicky1959} first proposed that multiple encounters in close vicinity might lead to total capture or disruption of galaxies. \cite{Alladin1965} showed the capture of spherical galaxies using hyperbolic orbits of encounter. \cite{Yabushita1971} and \cite{Tashpulatov1970} discussed  moderate hyperbolic and prolate ellipsoidal orbits for the motion of secondary companion in a merging pair, based on close encounters. \cite{TT1972} considered parabolic encounters to model the formation of bridge and tails in galaxy mergers.

Most of the simulations mentioned above were low resolution ones, and used only a few hundred particles to construct dynamical models of interacting galaxy pairs \citep{TT1972}. Further, they mostly modeled the galactic discs as self-gravitating stellar discs though the galactic disc is a gravitationally-coupled, multi-component system of stars, gas and the dark matter halo (see, for example, {\cite{Bodenheimer20017}).  Construction of a suitable dynamical model for an interacting pair of galaxies with observational constraints strongly hinges on the appropriate choice of initial conditions. The initial conditions may require specification of several dynamical parameters related to mass ratio, energy, spin and orbital geometry. In addition, another set of parameters like viewing directions, length and velocity scales have to be chosen to fit the model results with the observed structure and kinematics \citep{TT1972, galmer2010, Barnes2009, Barnes2011, Privon2013, Mortazavi2015}. As initial conditions are not known a priori, they are currently determined by trial and error method. However, this method is not practically feasible as the N-body + hydrodynamical simulations are computationally expensive.  As such, in our present work, our goal is to employ machine learning approaches to automatically determine initial conditions and other parameters for constructing dynamical models for interacting galaxy pairs.

Machine learning techniques have been applied to various problems in astronomy ranging from classifications of stars and galaxies \citep{Weir1995}, optical transients \citep{Cabrera-Vives2017, Mahabal2011}, galaxy morphology \citep{Dieleman2015}, radio galaxies \citep{Aniyan2017}, bar and unbarred galaxies \citep{Sheelu2018} to problems in photometric redshift \citep{Hoyle2016} and variable starlight \citep{Mahabal2017}. All these methods have been successful in terms of reliability, robustness, accuracy and computation. \citet{Aniyan2017} obtained 95\% precision for the detection of bent-tailed radio galaxies. Similarly, a precision of 94\% was achieved for the classification of barred and un-barred galaxies \citep{Sheelu2018}. Recently, \cite{Bekki2019} showed the application of Deep Convolutional Neural Networks (DCNN) on Smoothed Particle Hydrodynamics (SPH) simulation data to determine an optimal orbital geometry of satellite galaxies in galaxy clusters favourable for ram pressure stripping (RPS), and obtained an accuracy of 95\%. \\

We propose the use of Deep Convolutional Neural Networks (DCNNs) to determine parameters specifying initial conditions for a dynamical model of an interacting galaxy pair as well as viewing directions to match the model with observations. We use GalMer, an existing database of N-body + Smoothed Particle Hydrodynamical (SPH) simulations of galaxy mergers \citep{galmer2010} to train the CNN models. In this paper, we develop models to determine the relative inclination ($i$) between the discs and the viewing angle ($\theta$) which is the angle  between the line of sight and the normal to the orbital plane of an interacting pair of galaxies. Finally, we test our model on real images from SDSS DR15. To the best of our knowledge, this is the first instance of application of CNN to this problem (see, for example, \cite{MNRAS2020Blumenthal})  \\

The rest of the paper is organized as follows. In Section~\ref{sec:sec_2}, we describe the details of GalMer simulation, Section~\ref{dynparest} describes estimation of dynamical parameter using DCNN, in Section~\ref{sec:data_preprocessing} data collection and preprocessing. CNN architectures is introduced in Section~\ref{sec:sec_4}. We discuss experimental results in Section~\ref{sec:sec_5} followed by the conclusions in Section~\ref{sec:sec_6}.

\section{GalMer}
\label{sec:sec_2}
GalMer is a N-body + SPH galaxy merger simulation with moderate resolution. The galactic model chosen for simulation contains a spherical non-rotating dark-matter halo that may or may not contain a stellar disk, a gaseous disk and a central non-rotating bulge. The galaxies are represented in terms of 1,23,000 particles with total mass distributed among gas, stars and dark matter. It also allows direct comparison between simulation and observation using Virtual Observatory (VO) tools \footnote{\href{http://www.projet-horizon.fr}{http://www.projet-horizon.fr}} framework. 

GalMer covers interactions between galaxies of different mass ratios, morphological types, orbit-types, spin types
and finally different values of the relative inclination $i$ between the discs. GalMer VO tool allows users to choose different values of the above parameters. The different morphological types included were a giant galaxy (giant elliptical ($gE_o$), giant lenticular ($gS_o$), giant spiral ($gS_b$, $gS_a$, $gS_c$, $gS_d$)) with another giant type ($gE_o$, $gS_o$, $gS_b$, $gS_a$, $gS_c$, $gS_d$) or intermediate-mass ($iE_o$, $iS_o$, $iS_a$, $iS_b$, $iS_d$) or dwarf galaxy ($dE_o$, $dS_o$, $dS_a$, $dS_b$, $dS_d$). As discussed in \citep{galmer2010}, GalMer database consists of interacting galaxy pairs with mass-ratios, 1:1, 1:2 and 1:10, corresponding to giant-giant, giant-intermediate and giant-dwarf combinations, respectively. Based on different values of initial distance, pericentric distance and motional energies, 12 different orbit types were chosen. The spin types considered were prograde or retrograde. For the 1:1 mass ratio case, the relative inclination $i$ between the discs of the interacting galaxies was fixed at any one of the following four values: $0^{\circ}, 45^{\circ}, 75^{\circ} \text{ and } 90^{\circ}$. Similarly, for the 1:2 and 1:10 mass ratio cases, $i$ was fixed at 33$^{\circ}$ only. We may note here that $i$ is a continuous variable and may assume any value between $0^{\circ}$ and $90^{\circ}$. However, the above set of simulations for a few discrete values of $i$, roughly spanning its allowed range of values, is good enough to fairly understand the role of $i$ in regulating interacting galaxy morphology. Finally, in order to compare the simulation results with observations, one can also change (i) angle between the normal to the orbital plane and the line of sight, $\theta$, between $0^{\circ} - 90^{\circ}$ and (ii) the azimuthal angle, $\phi$, subtended by the long axis of the galaxy between $0^{\circ} - 360^{\circ}$.

In Fig.~\ref{fig:geometry}, we represent a schematic of interaction between two galaxies. In this figure, the interacting pair of galaxies are denoted by $P_1$ and $P_2$. The normal to the orbital plane is taken to coincide with the z-axis of the 3-D cartesian coordinate system. The angle between the spin axis and the axis of orbital motion i.e., $z$ axis are given by $i_1$ and $i_2$ for $P_1$ and $P_2$ respectively. In addition, azimuthal angles for both the galaxies, ${\Phi}_1$ and ${\Phi}_2$, are taken to be zero. In all GalMer simulations, the first galaxy of the interacting pair is chosen to lie on the orbital plane i.e., $i_1$ = 0.  Therefore, the relative inclination between the discs of an interacting pair of galaxy in GalMer simulation is given by $i = i_2 - i_1 = i_2$.  The probability of orientation of $P_2$ between 0 and $i_2$ is given by $1 - \rm{cos(i_2)}$. The viewing angle which is the angle made by line of sight of an observer with normal to the orbital plane,  is denoted by $\theta$. \\

\begin{figure}
    \centering
    \includegraphics[width = \columnwidth]{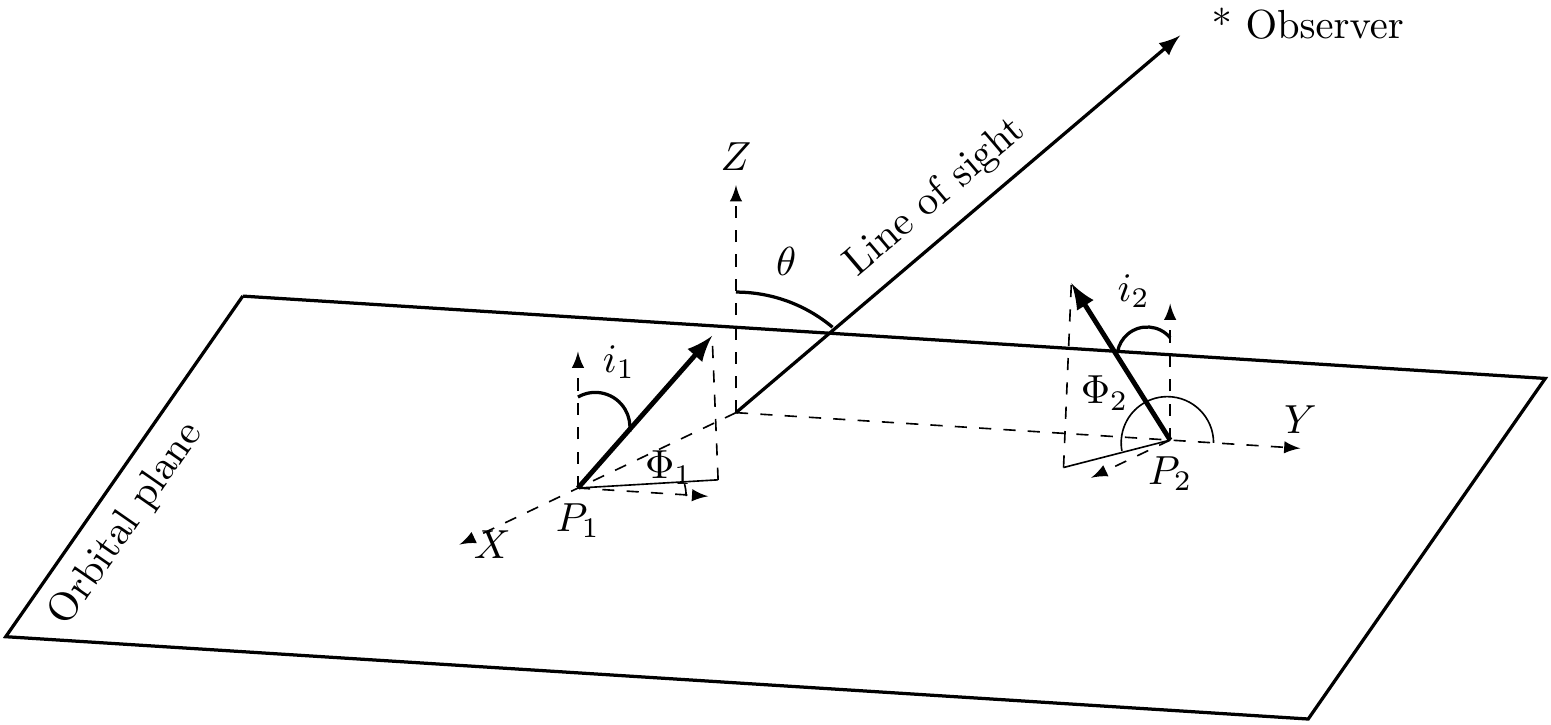}
    \caption{3D Cartesian coordinate system of GalMer. Inclinations ($i_1$ and $i_2$) of the galaxies ($P_1$ and $P_2$) are measured with respect to z-axis. $\Phi_1$ and $\Phi_2$ are the corresponding azimuthal angles of galaxies with respect to y-axis and $\theta$ is the viewing angle. Relative inclination is given by, $i = i_2 - i_1$. For GalMer, $i_1 = 0$, $\Phi_1 = \Phi_2 = 0$.}
    \label{fig:geometry}
\end{figure}

Each galaxy interaction includes 50 to 70 snapshots at an interval of 50 Myr starting from 0 to 2.5/3.5 Gyr. The meta-data produced during interaction can be visualized through the web interface.

\section{Dynamical parameter estimation using DCNN}
\label{dynparest}
As mentioned earlier, in our present work, we propose to use DCNN for the estimation of dynamical parameters like relative inclination ($i$) and viewing angle ($\theta$). Theoretically, these angles can have continuous values in the range $[0^{\circ}, 90^{\circ}]$ and $[-90^{\circ}, 90^{\circ}]$ respectively. But GalMer simulates only a discrete set of $i$ values. For example, ($0^{\circ}, 45^{\circ}, 75^{\circ} \text{ and } 90^{\circ}$) for the 1:1 mass ratio case. But for each $i$, almost all possible values of $\theta$ values in the range, $[-90^{\circ}, 90^{\circ}]$, can be simulated.

For any supervised machine learning problem, number of classes can be finite or infinite. If the number of target classes is finite or discrete, it is considered as a classification problem. Otherwise, if the number of classes is infinite or continuous, the problem is considered as a regression problem \citep{duda2012pattern}. Due to the availability of simulations corresponding to only a very few $i$ values in GalMer database, we have used DCNN based classification to determine these values. If simulations corresponding to many different values of $i$ within the allowed range (thus mimicking an almost continuous range of $i$) were available, regression could be used. Similarly,  as simulations corresponding to an almost continuous range of $\theta$ values were available, DCNN based regression has been used for their estimation.

Deep Convolutional Neural Network requires a huge amount of data for training to achieve a desirable accuracy \citep{duda2012pattern}. Training data should also be consistent, in the sense, it should represent a single phenomenon or pattern. For example, consider image data samples of galaxy pairs with same relative inclination but with different mass ratios. Due to different ratios, there may be difference in sizes/appearance of galaxies in the interacting pair. As DCNN looks and extracts features only based on image pixels, there is some tolerance on the ratio, beyond which features learnt may be changed for the same relative inclination angle, that is not intuitive. That means, DCNN can be applied on samples from a range of mass ratios that generate consistent image patterns. As GalMer supports only three ratios (1:1, 1:2 and 1:10), we have chosen mass-ratio of 1:1, in our experimentation, for the sake of consistency. Accordingly, we have considered interactions between morphological types $gS_a$ and $gS_b$ only i.e., $gS_a - gS_a$, $gS_a - gS_b$ and $gS_b - gS_b$) at their pericentric approach.

Based on GalMer environment, we have performed a series of experiments using a single parameter ($i$ or $\theta$) as well as both the parameters ($i$ and $\theta$). As discussed above, we have used classification for $i$ parameter and regression for $\theta$. As $i$ is discrete, we have used classification for the combination of both the parameters ($i$ and $\theta$).

We first attempt a 2-class classification based on only one parameter, $i$: $i = 0^{\circ}, 45^{\circ}$. We then implement a 3-class classification with $i = 0^{\circ}, 45^{\circ}, 90^{\circ}$. For each value of $i$ in the above two classification cases, we consider samples with different values of $\theta$: $0^{\circ}, 20^{\circ}, 35^{\circ}, 50^{\circ}, 65^{\circ}, 80^{\circ}$. We have also performed classification experiments based on both $i$ and $\theta$ parameters. As $i$ is discrete in GalMer, we have chosen some discrete values for $\theta$ also, as follows: ($(i, \theta) \sim$ $(0^{\circ}, 15^{\circ})$, $(0^{\circ}, 45^{\circ})$, $(0^{\circ}, 90^{\circ})$, $(45^{\circ}, 15^{\circ})$, $(45^{\circ}, 45^{\circ})$, $(45^{\circ}, 90^{\circ})$, $(90^{\circ}, 15^{\circ})$, $(90^{\circ}, 45^{\circ})$, $(90^{\circ}, 90^{\circ})$). In this case, for each $\theta$, we actually considered a range of values centered around the quoted $\theta$ value, $[\theta-5^\circ$, $\theta+5^\circ]$, to augment the dataset. 

To estimate $\theta$ value alone, we have used CNN based regression \citep{deeplearnebook}. For a given $i$ value,  image samples for different values of $\theta$ in the range, $[-90^{\circ}, 90^{\circ}]$, have been used for regression.

Though there are around 9 dynamical parameters, we have focused only on relative inclination $i$ and viewing angle $\theta$ in the current work. Also, we have not aimed to classify the images based on additional parameters closely associated with relative inclination like a prograde/retrograde passage: A prograde passage is one in which the direction of the galactic spin is aligned with that of the orbital motion whereas a retrograde passage is one in which the former oppositely aligned with the latter. A prograde passage will have a different dynamical signature as compared to a retrograde passage. However, in this debut attempt, we developed a basic CNN to determine two parameters, $i$ and $\theta$, only. So, for the determination of $i$, we have combined samples for both the spin directions and several orbit types. However, our CNN framework can be easily extended to consider additional dynamical parameters like prograde/retrograde passage. For example, for a given $i$ value, we can also collect samples for each spin direction and train the CNN by just increasing the number of target classes to two times (one for each direction like $0^\circ$ Prograde, $0^\circ$ Retrograde, $45^\circ$ Prograde, $45^\circ$ Retrograde etc.)

\subsection*{Significance of $i$ and $\theta$ in dynamical models of interacting galaxy pairs}

In general, $i$ and $\theta$ together determine the overall geometry of the interacting galaxy pair as projected in the sky.  
$i$ is the relative inclination between the discs of the interacting galaxy pair. In the GalMer framework, $i$ is also the inclination of the second galaxy in the pair with respect to the orbital plane as the first galaxy is taken to be aligned with the orbital plane (See Section~\ref{sec:data_preprocessing}). Therefore $i$ is a dynamical parameter and is therefore expected to regulate the structure and dynamics of the system. Using simplistic test particle simulations, \citet{TT1972} showed that the value of $i$, as understood in the GalMer framework, strongly governs the formation of tidal features in interacting galaxies. For an unequal mass encounter, a smaller value of $i$ in a prograde passage results in the formation of longer tidal bridges.
Similarly, for an equal mass encounter, a smaller value of $i$ in a prograde passage leads to the formation of long and curved tails. In a retrograde passage, on the other hand, tidal features may not be well developed. The above simple models already hint at the dynamical significance of the value of $i$ in regulating the detailed morphology of an interacting galaxy pair. However, the results may deviate a bit in a real scenario where the galactic discs are self-gravitating. In general, however, the relative inclination between the galaxies of an interacting pair is not the same as the relative inclination with respect to the orbital plane. Besides, the observed images do not directly indicate the inclination of each galaxy with respect to the orbital plane but rather the inclination with the plane of the sky, i.e., the plane perpendicular to the line of sight. While the latter mainly determines the overall geometry of the interacting system, the former regulates the details of the galaxy morphology. As discussed above, perfect alignment with the orbital plane will result in the development of distinct tidal features as opposed to the case in which the plane of the galaxy is perpendicular to the orbital plane. In other words, the details of this galaxy morphology as revealed in the observed images may be used as a diagnostic tracer of the relative angle of inclination with the orbital plane of the galaxy.

In the GalMer framework, $\theta$ is the angle between the line of sight and the normal to the orbital plane, which also happens to be the plane of the first galaxy of the interacting pair (See Section~\ref{sec:sec_2}). It is an observational parameter used to project the simulated image from the orbital plane to the sky plane to match the observed images and therefore does not regulate the dynamics of the system. However, as discussed above, it does regulate the geometry of the interacting galaxy pair as projected in the sky. For example, the length of tidal features may appear very different when projected in the sky plane. Therefore $\theta$ plays an essential role in determining the correct dynamical model for the observed images of interacting galaxy pair.

Therefore, in our first attempt to use CNN in determining the dynamical and observational parameters characterizing an interacting pair of galaxies, we choose to determine $i$ and $\theta$ only. However, we plan to extend the model to include the other parameters, which will crucially hinge on the availability of enough simulation images, spanning the multi-dimensional dynamical and observational parameter space of an interacting pair of galaxies.

\section{Data collection and Preprocessing}
\label{sec:data_preprocessing}

We have initially downloaded images of size $450 \times 450$ in GIF (Graphics Interchange Format) format from the GalMer website. The total number of images obtained from the above website is only a few hundred in number, which may not be sufficient to train a Deep CNN. As such, we have applied some standard Image Processing operations like smoothing, contrast enhancement,  etc., on the initial set of images. For contrast enhancement, we have randomly selected one half of the image, either along the horizontal or along the vertical direction, and then changed the contrast of the same with respect to the rest of the image. Finally, all the above augmentation operations resulted in an increase in the size of the image dataset by a factor of 3.

As mentioned in \citep{krizhevsky2012}, we have further augmented the database by rotating all the images, by one degree, 359 times. 

After applying the augmentation steps, all the images are scaled to $340 \times 340$ size, to reduce computational complexity. In the augmented database, we have used $80\%$ of the total number of images for training and validation, and the remaining $20\%$ for testing. Validation set can be used to stop the learning process if the training leads to overfitting \citep{krizhevsky2012}. We have shown the distribution of data into training, validation and testing sets, for two, three and nine classes, in  Tables~\ref{tab:tab_0_45}, \ref{tab:tab_0_45_90}, and \ref{tab:tab_data_incl_full} respectively. In Table~\ref{tab:tab_data_incl_full} for nine classes, as discussed earlier, each class is denoted by a pair of angles in which first one gives relative inclination $i$ of an interacting pair of galaxies while the second one  gives the viewing angle $\theta$. For example, in the first table entry, $(0^\circ, 15^\circ)$, $i$ and $\theta$ values are $\mathrm{0^\circ}$ and $\mathrm{15^\circ}$ respectively. 

\setlength{\arrayrulewidth}{0.2mm}
\renewcommand{\arraystretch}{1.0}

 \begin{table}
 \centering
 \caption{Distribution of data for two class ($0^\circ$ and $45^\circ$) classification}
 \label{tab:tab_0_45}
 \begin{tabular}{>{\bfseries}lccc }
 \hline
 \multicolumn{4}{c}{\textbf{Data Set}} \\
 \hline
 \textbf{Class}   &  \textbf{TRAINING} & \textbf{ VALIDATION} & \textbf{TESTING}\\
 \hline
  $\mathrm{0^o}$ & 5,58,316 & 1,41,524 & 486 \\

  $\mathrm{45^o}$ & 5,23,733   & 1,32,907 & 456\\
 
 \hline
 \end{tabular}
\end{table}

\begin{table}
 \centering
 \caption{Distribution of data for three class ($0^\circ$, $45^\circ$ and $90^\circ$) classification}
 \label{tab:tab_0_45_90}
 \begin{tabular}{>{\bfseries}lccc }
 \hline
 \multicolumn{4}{c}{\textbf{Data Set}} \\
 \hline
 \textbf{Class}   &  \textbf{TRAINING} & \textbf{ VALIDATION} & \textbf{TESTING} \\
 \hline
  $\mathrm{0^o}$ & 5,58,316 & 1,41,524 & 486 \\
 
 $\mathrm{45^o}$ & 5,23,733   & 1,32,907 & 446 \\
 
 $\mathrm{90^o}$ & 6,22,480  &  1,55,120 & 540 \\
 \hline
 \end{tabular}
\end{table}

\begin{table}
 \centering
 \caption{Distribution of data for multi-class classification. Here, each class is denoted by a pair of relative inclination and viewing angles.}
 \label{tab:tab_data_incl_full}
 \begin{tabular}{>{\bfseries}lccc }
 \hline
 \multicolumn{4}{c}{\textbf{Data Set}} \\
 \hline
 \textbf{Class}   &  \textbf{TRAINING} & \textbf{VALIDATION} & \textbf{TESTING} \\
 \hline
 $\mathrm{(0^o,15^o)}$ & 3,42,055 & 85,514 &  510 \\
 
 $\mathrm{(0^o,45^o)}$ & 3,42,055  & 85,514 & 510 \\
 
 $\mathrm{(0^o,75^o)}$ & 3,42,055 & 85,514 &  510 \\
 
 $\mathrm{(45^o,15^o)}$ & 3,03,858 & 75,964 &  454 \\
 
 $\mathrm{(45^o,45^o)}$ & 3,03,858 & 75,964 &  454 \\
 
 $\mathrm{(45^o,75^0)}$ & 3,03,858 & 75,964 &  454 \\
 
 $\mathrm{(90^o,15^o)}$ & 3,71,349 & 92,838 &  555 \\
 
 $\mathrm{(90^o,45^o)}$ & 3,71,349 & 92,838 &  555 \\
 
 $\mathrm{(90^o,75^o)}$& 3,71,349 & 92,838 &  555 \\
 \hline
 \end{tabular}
\end{table}

\section{Convolutional Neural Networks} \label{sec:sec_4}
Convolutional Neural Networks, can be considered as an extension of Artificial Neural Networks (ANNs) to handle 2D data like images. However, the underlying principle for learning is the same, based on error backpropagation \citep{rumelhart1986learning, hecht1989neurocomputer} which minimizes the error function, $e$, shown in equation~\ref{eq:eq_3}. Here, $y$ is the obtained output of the neuron and $\hat{y}$ is the actual output.

\begin{equation}
    \label{eq:eq_3}
    \mathrm{
    e = \frac{1}{2} ||\hat{y} - y ||^{2}}
\end{equation}

Initially, CNN has been proposed for handwritten digit recognition by \citep{lecun2015}. Later it has been applied to other types of problems like regression \citep{hagenauer1996iterative} and also on various kinds of data like text, video, speech etc. CNN facilitates automatic feature extraction using a sequence of convolution and pooling layers. Low-level to high-level features are learnt from the input data, in consecutive layers of the network, from left to right \citep{zeiler2014visualizing, Oquab_2014_CVPR, krizhevsky2012}. In the final fully connected output layer, a loss function \citep{hagenauer1996iterative} like mean squared error, cross entropy etc., is used to optimize parameters of the network. 

In our present work, we have used Deep CNN \citep{krizhevsky2012, yosinski2014transferable} for classification and regression of dynamical parameters using GalMer's synthesized images. In the literature, DCNN has been widely used in the field of astronomy for different problems like bar detection in galaxies \citep{Sheelu2018,Dieleman2015}, classification of radio galaxies \citep{Aniyan2017}, optical transients \citep{Cabrera-Vives2017}, photometric redshifts \citep{Hoyle2016}, variable stars \citep{Mahabal2017}, RPS models \citep{Bekki2019} etc. 

\subsection{Network Design}
A CNN based model is configurable based on tunable hyper parameters like number of layers and feature maps, kernel size, learning rate, activation functions etc., and the network has to be optimized for a given problem and domain, based on these parameters. As our primary focus is on the determination of dynamical parameters, we have used existing popular architectures like AlexNet, instead of proposing and optimizing any new architectures. In the literature \citep{krizhevsky2012}, it is mentioned that design issues for some model architectures like AlexNet have been extensively explored and optimized for a large number of classes and datasets, and so we have chosen this architecture. These architectures are also made publicly available.

AlexNet design framework, which is shown in Table~\ref{tab:tab_layer}, contains a total of 12 layers including Convolutional (Conv), Max pooling (Pool) and fully connected (FC) layers. Max pooling layers are used to down-sample the input size to reduce the computational complexity. The model also contains Dropout layers \citep{srivastava14a} with 50 percent dropout rates to remove weakly connected neurons in the FC layers. All the convolutional layers are followed by Rectified Linear Unit activation (ReLU) \citep{relu2010} and Normalisation (Norm) functions. The last pooling layer is  connected to a series of FC layers followed by a final Softmax layer \citep{gold1996}. These FC layers also use ReLU activation function and the Softmax layer determines the prediction probabilities of the input image in various classes. Input image is classified to a class with highest prediction probability. 

As mentioned earlier, CNN can also be used for regression \citep{deeplearnebook} by using a fully connected single node output layer with a linear activation function, instead of softmax classifier output layer. The network has to be trained using continuous prediction loss functions such as mean squared error, mean absolute error, mean absolute percentage error, etc. The goal of regression is to fit a function over a set of training points (in our case, each point is given by, image and $\theta$) by minimizing a loss function. In our experimentation, we have used mean squared error (MSE) as a loss function, which is given by equation \ref{eq:msefunc}.  Here, $n$ is the number of points used for fitting a function, $f$, using regression. $f_i$ is the estimated or predicted value for point $i$ and $y_i$ is the actual value. Ideally, MSE is zero for the perfect fit.

\begin{equation}
    \label{eq:msefunc}
    \mathrm{
    \text{MSE} = \frac{1}{n} \sum_{i=1}^n (y_i-f_i)^2}
\end{equation}

As we have limited number of samples for $\theta$ values in the range, $[-90, 90]$, we have not used AlexNet architecture. As shown in Table \ref{tab:tab_regress}, we have used architecture with 3 convolutional and pooling layers with less number of feature maps in each layer, followed by 3 fully connected layers. As discussed above, last fully connected layer with single node gives the continuous estimated output value using the fitted function. These type of architectures have been used for deep learning based Computer Vision problems \citep{deeplearnebook}. However, we have optimized the network with respect to hyperparameters like optimization algorithm (Eg: Adam, Stochastic Gradient Descent etc.), learning and decay rates etc.

\begin{table}
 \centering
 \caption{AlexNet architecture for classification}
 \label{tab:tab_layer}
 \scalebox{0.9}{
 \begin{tabular}{>{\bfseries}ccccc}
 \hline
 & & & & \\
 \textbf{\parbox{1cm}{Layer number}} & \textbf{Layer type}   &  \textbf{\parbox{1.5cm}{ No. of Kernels/Pooling windows} } & \textbf{\parbox{1.5cm}{Kernel/ Pooling window size}} & \textbf{\parbox{1.5cm}{ Parameters}}\\
 & & & & \\
 \hline
 \hline
 1 & Convolutional & 96 & 11 x 11 & 34944 \\
  \hline
 2 & Pooling & 96 & 3 x 3  & 34944\\
 \hline
 3 & Convolutional & 256 & 5 x 5 & 614656 \\ 
\hline
 4 & Pooling & 256 & 3 x 3 & 614656\\
 \hline
 5 & Convolutional  & 384 & 3 x 3 & 885120 \\ 
 \hline
 6 & Convolutional & 384 & 3 x 3 & 1327488 \\
 \hline
 7 & Convolutional & 256 & 3 x 3 & 884992 \\
 \hline
 8 & Pooling & 256 & 3 x 3 & 884992 \\
 \hline
 9 & Fully Connected &  4096 & NA & 4198400 \\
 \hline
 10 & Fully Connected &  4096 & NA & 16781312 \\
 \hline
 11 & Fully Connected & 4096 x 2 & NA & 8194 \\
 \hline
 \hline
 12 & Softmax &  \multicolumn{3}{l}{2/3/9 (based \par on number of classes)} \\
 \hline
 \end{tabular}}
\end{table}

\begin{table}
 \centering
 \caption{CNN architecture for regression}
 \label{tab:tab_regress}
 \scalebox{0.9}{
 \begin{tabular}{>{\bfseries}ccccc}
 \hline
 & & & & \\
 \textbf{\parbox{1cm}{Layer number}} & \textbf{Layer type}   &  \textbf{\parbox{1.5cm}{ No. of Kernels/Pooling windows} } & \textbf{\parbox{1.5cm}{Kernel/ Pooling window size}} & \textbf{\parbox{1.5cm}{ Parameters}}\\
 & & & & \\
 \hline
 \hline 
 1 & Convolutional & 16 & $3 \times 3$ & 448 \\ \hline
 2 & Pooling & 16 & $2 \times 2$ & 448 \\ \hline
 3 & Convolutional & 32 & $3 \times 3$ & 4640 \\ \hline
 4 & Pooling & 32 & $2 \times 2$ & 4640 \\ \hline
 5 & Convolutional & 64 & $3 \times 3$ & 18496 \\ \hline
 6 & Pooling & 64 & $2 \times 2$ & 18496 \\ \hline
 7 & Fully Connected & 16 & NA & 1048592 \\ \hline
 8 & Fully Connected & 4 & NA & 68 \\ \hline\hline
 9 & Fully Connected & 1 & NA & 5 \\ \hline
 \end{tabular}}
\end{table}

\subsection{Automated feature extraction}
As discussed earlier, CNN automatically extracts low to high level features using kernel windows, directly from the input images. In Fig.~\ref{fig:filters}(a)-(f), we have shown an input image and some of its feature maps, extracted using one kernel window of each convolutional layer from left to right. Low level features are learnt in the lower layers and as we progress to higher layers from left to right, high level abstract features are computed by the fusion of lower level features. In the given figure, it can be seen that shapes of galaxies are captured after the first layer and their galactic centers in the second layer. Complex abstract features like line segment joining the centers, the corresponding slope/inclination etc., are learnt automatically in the higher layers from the combination of feature maps in the previous layers. These abstract features are ultimately used for the final classification task.

\begin{figure*}
    \centering
    \includegraphics[width=16cm]{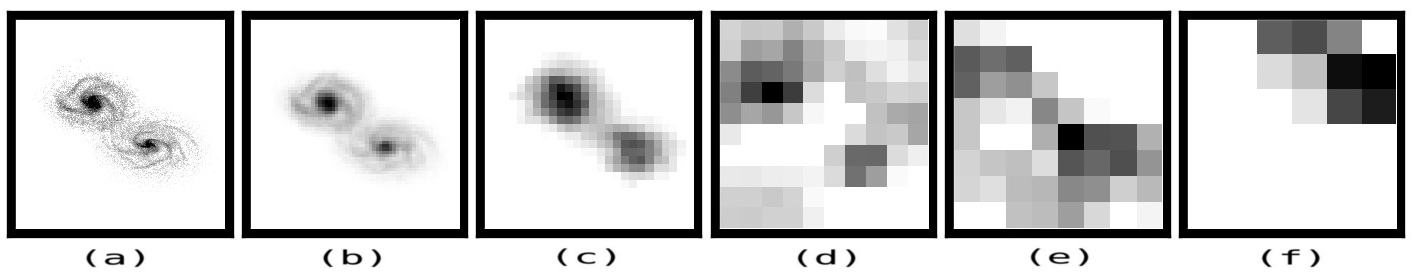}
    \caption{CNN automated feature extraction. (a) Input Image (b)-(f) Feature maps obtained for (a) using one kernel window from each of the five convolutional layers}
    \label{fig:filters}
\end{figure*}

\section{Experimental results and Discussion} \label{sec:sec_5}
We have implemented the AlexNet model in Python using the publicly available Keras library \citep{chollet2015keras} for deep learning. As training involves heavy computations, we have used Intel(R) Core i7-7700 processor with 3.60GHz frequency and 16 GB RAM. As mentioned earlier, we have performed classification experiments using 2 and 3 classes based on $i$ alone, and 9 classes based on both $i$ and $\theta$. We have also performed regression experiment based on $\theta$ alone. Input images are scaled down to $256 \times 256$ size for all the experiments. We have discussed each of these four experiments in the following subsections.

For classification experiments (first 3 cases below), training process takes an average time of  around 8 to 10 hours using stochastic gradient algorithm with a learning rate of 0.01 and a decay rate of 1\%, for running one full epoch. We have used different number of epochs and steps per epoch for different number of classes, discussed earlier. For two classes, training has been performed with 12 epochs and 12000 steps per epoch. Similarly, 14 epochs with 12000 steps per epoch, and 44 epochs with 16000 steps per epoch, have been used for 3 and 9 classes respectively. We have also presented learning curves that show the progress of training process, for the above three different cases in Figure 3. In these figure,  training accuracy and validation loss functions with respect to number of epochs are plotted. It can be observed that the model has achieved an accuracy of 99.9\% just after 3 epochs in the case of 2-class classification. If the number of classes is increased to 3 and 9, more number of epochs is required to achieve the same accuracy. Montages of some sample images for different cases of classification are shown in Appendix \ref{append1}.

\begin{figure*}
\begin{center}
\begin{tabular}{ccc}
\resizebox{55mm}{55mm}{\includegraphics{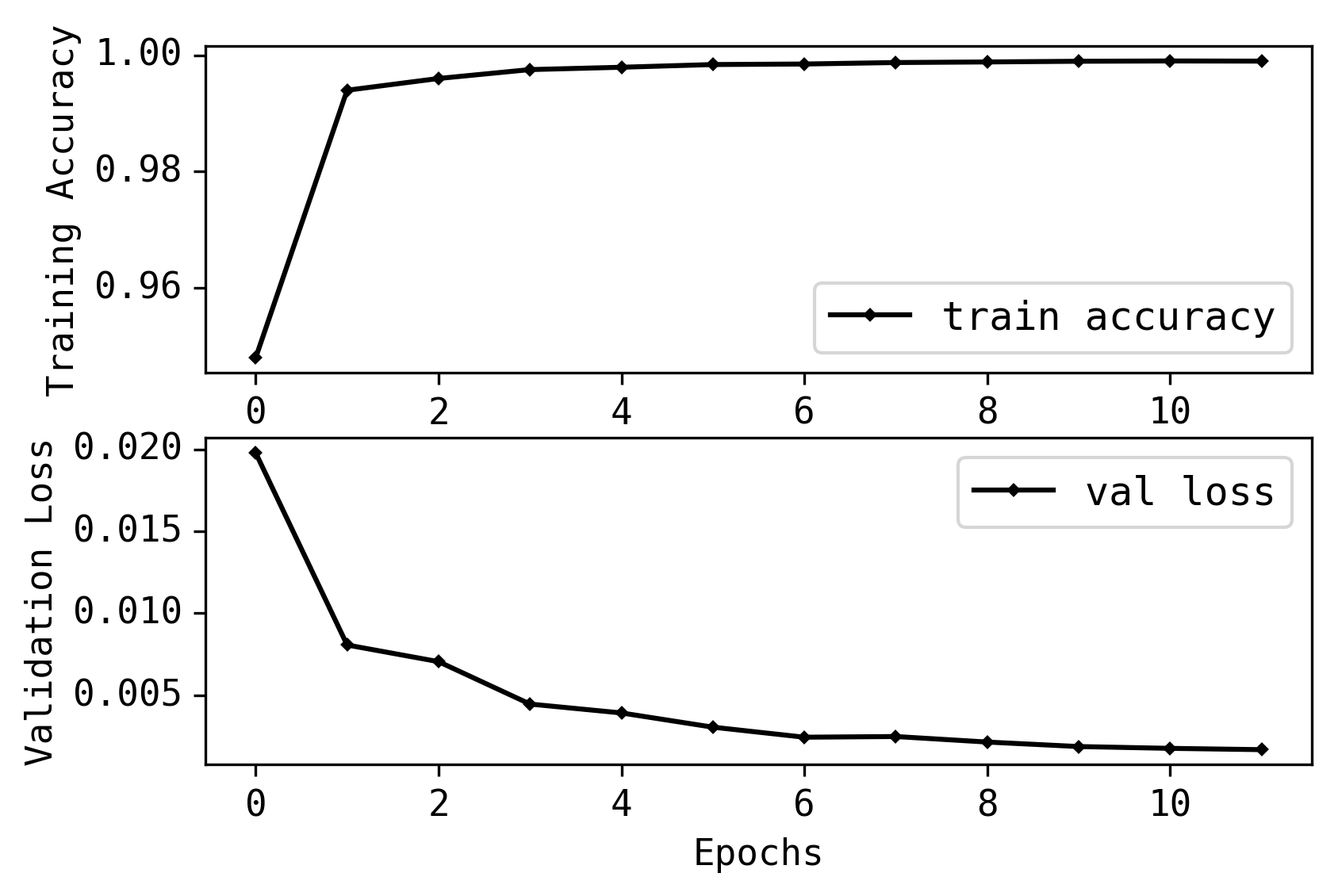}} &
\resizebox{55mm}{55mm}{\includegraphics{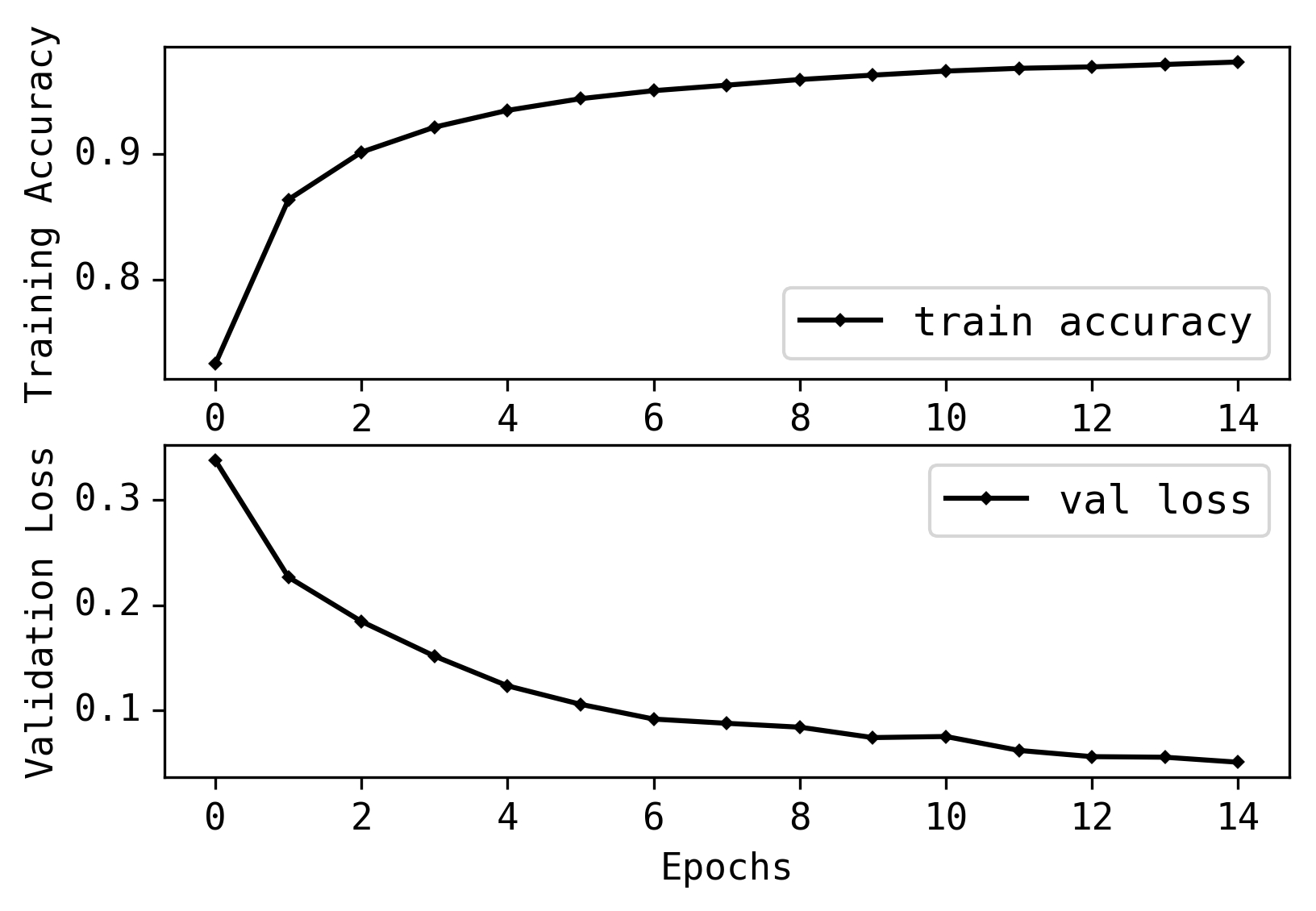}} &
\resizebox{55mm}{55mm}{\includegraphics{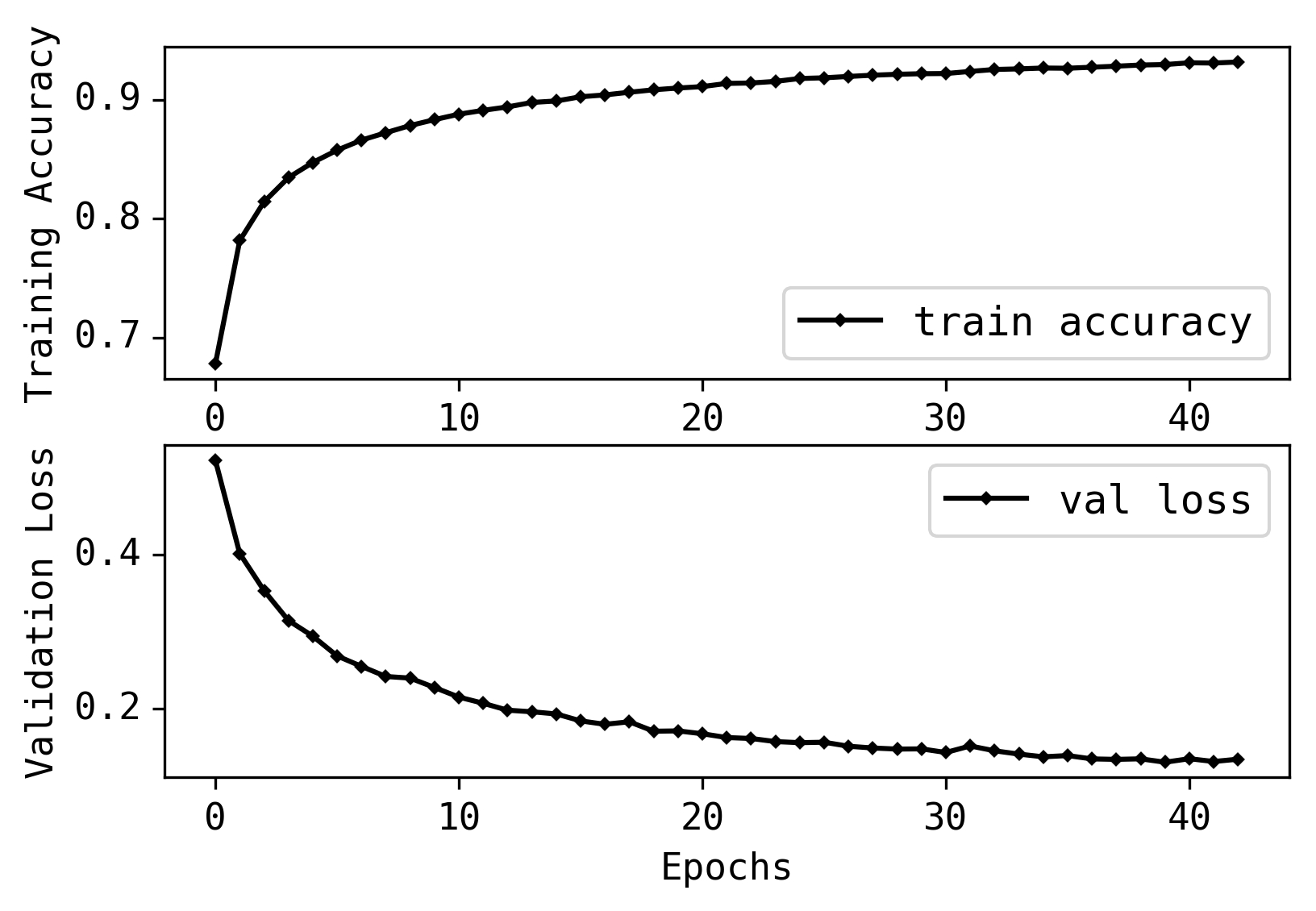}}\\
\end{tabular}
\end{center}
\caption{Left Panel: Learning curve for two classes, 0 and 45 degrees. Middle Panel: Learning curve for three classes, 0, 45 and 90 degrees 
Right Panel: Learning curve for 9 classes (pair of angles)}
\end{figure*}

After training the model, we have evaluated the network performance using test data. Performance has been quantified in terms of standard precision, recall and $F$ measures \cite{Sheelu2018}, whose expressions are discussed below.  

\begin{equation}
    \mathrm{Precison = \frac{TP}{TP + FP}}
\end{equation}

\begin{equation}
    \mathrm{Recall = \frac{TP}{TP+FN}}
\end{equation}

\begin{equation}
    \mathrm{F_{\beta} = (1 + \beta^{2}) \cdot \frac{Precision x Recall}{Precision + Recall}}
\end{equation}

Here, TP denotes True Positive which is given by the number of correctly classified images in a class. FP and FN denote False Positive and False Negative, which is given by the number of incorrectly classified images (images from first class are classified as second one and vice-versa). In $F$ measure, $\beta$ determines the weightage of Precision and Recall values. We have set $\beta = 1$ and used $F_1$ score to give equal weightage for both the values. 

\subsection*{Case 1: 2-class classification}
\label{sec:sec 5.1}
In Table~\ref{tab:tab_0_45_PR}, we have shown Precision, Recall and $F_{1}$ scores for 2-class classification. If true positive and false positive rates are plotted, as a point on Receiver Operating Characteristic (ROC) curve, the area under the curve (AUC) is 0.96. This value is closer to 1, which means that the model has a very good prediction capability. 
\begin{table}
 \centering
 \caption{Precision,recall and $F$ measures for 2-class classification, 0 and 45 degrees.}
 \label{tab:tab_0_45_PR}
 \begin{tabular}{ >{\bfseries}lcccc }
 \hline
     &  \textbf{Precision (\%)}  &  \textbf{Recall (\%)} & \textbf{$F_{1}$-Score (\%)} & \textbf{Total} \\
 \hline
 $\mathrm{0^o}$ & 0.98 & 1.00 & 0.99 & 478\\

 $\mathrm{45^o}$ & 1.00 & 0.97 & 0.99 &  439\\
 \hline
 Overall &  0.99 & 0.99 & 0.99 & 917\\
 \hline
 \end{tabular}
\end{table} 

 We have also considered rejections in our performance evaluation. If the highest probability in the Softmax layer, is less than some threshold, which is rejection threshold, the corresponding input image has been rejected. That means, network has not predicted the image with high confidence as determined by the rejection threshold. In our experimentation, we have chosen 80\% as the value of rejection threshold.  In Table \ref{tab:tab_0_45_rej}, number of rejections for each class are shown. In Fig~\ref{fig:conf_mat_0_45}, we have presented confusion matrix that gives statistics on true positives, false positives, true negatives and false negatives. Confusion matrices have been computed excluding the above mentioned rejections. 
\begin{table}
 \centering
 \caption{Number of rejected images from each class in 2-class classification. Rejection threshold is 80\%}
 \label{tab:tab_0_45_rej}
 \begin{tabular}{ >{\bfseries}lc }
 \hline
    Class &  \textbf{ Rejection count} \\
 \hline
 $\mathrm{0^o}$ & 8 \\

 $\mathrm{45^o}$ & 17\\
 \hline
 \end{tabular}
\end{table}

\begin{figure}
 \includegraphics[width=\columnwidth]{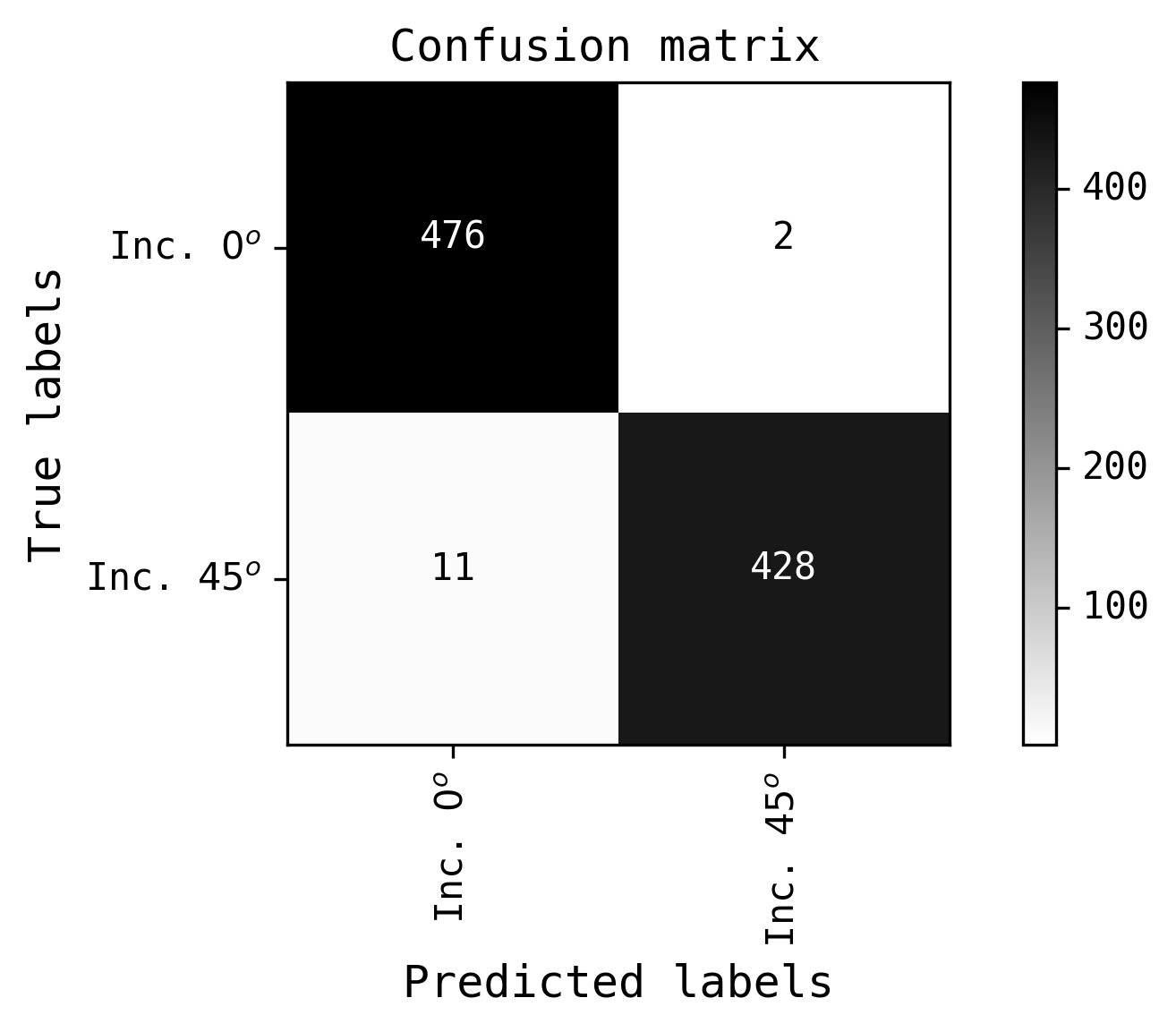}
 \caption{Confusion matrix for 2-class classification, 0 and 45 degrees.}
 \label{fig:conf_mat_0_45}
\end{figure}

\subsection*{Case 2: 3-class classification}
\label{sec:sec 5.2}
For this case, precision, recall and $F_{1}$ scores are shown in Table \ref{tab:tab_0_45_90_PR}. We have not plotted ROC curve as it is not straightforward for multi-class classification. Rejected image statistics are shown in Table \ref{tab:tab_0_45_90_rej} and the confusion matrix is presented in Fig. \ref{fig:conf_mat_0_45_90}. 

\begin{table}
 \centering
 \caption{Precision, recall and $F$ measures for 3-class classification, 0, 45 and 90 degrees.}
 \label{tab:tab_0_45_90_PR}
 \begin{tabular}{>{\bfseries}lcccc }
 \hline
   &  \textbf{Precision (\%)}  &  \textbf{Recall (\%)} & \textbf{$F_{1}$-Score (\%)} & \textbf{Total} \\
 \hline
 $\mathrm{0^o}$ & 0.99 & 0.99 & 0.99 & 474\\
 
 $\mathrm{45^o}$  & 0.98   & 0.98 & 0.98 &  410\\
 
 $\mathrm{90^o}$ & 0.97 & 0.97 & 0.97 & 493 \\
 \hline
 Overall &  0.98 & 0.98 & 0.98 & 1377\\
 \hline
 \end{tabular}
\end{table}

\begin{table}
 \centering
 \caption{Number of rejected images from each class in 3-class classification. Rejection threshold is 80\%}
 \label{tab:tab_0_45_90_rej}
 \begin{tabular}{>{\bfseries}lc }
 \hline
 Class  &  \textbf{Rejection count}  \\
 \hline
 $\mathrm{0^o}$ & 12 \\
 
 $\mathrm{45^o}$  & 46 \\
 
 $\mathrm{90^o}$ & 47 \\
 \hline
 \end{tabular}
\end{table}

\begin{figure}
 \includegraphics[width=\columnwidth]{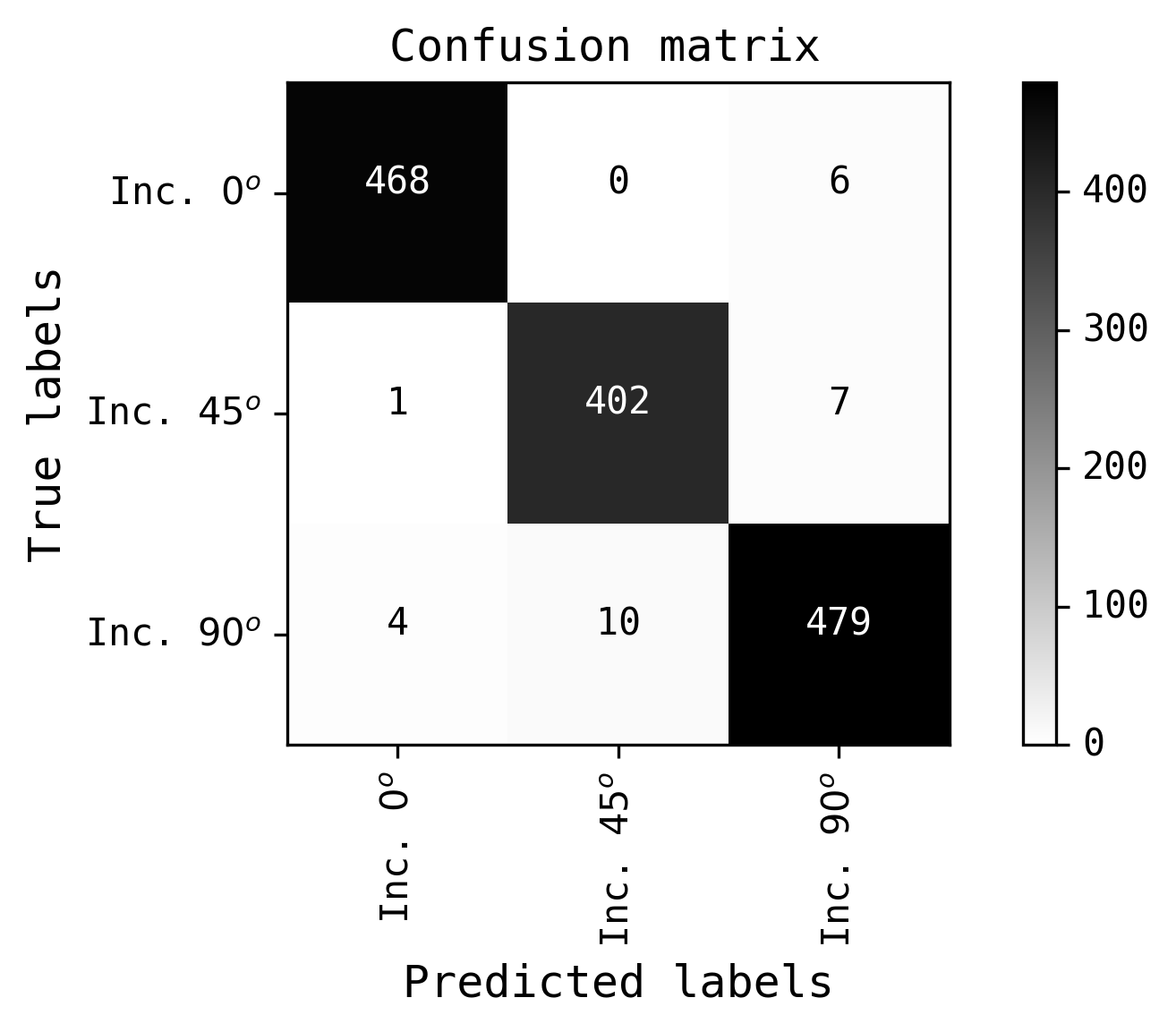}
 \caption{Confusion matrix for 3-class classification, 0, 45 and 90 degrees.}
 \label{fig:conf_mat_0_45_90}
\end{figure}

\subsection*{Case 3: 9-class classification}
\label{sec:sec 5.3}
As mentioned earlier, in this case, each class denotes a pair of relative inclination and viewing angles. As the number of classes is increased, prediction accuracy tends to reduce, which can be seen from the values of performance measures in Table~\ref{tab:tab_incl_PR}. This happens if the data may  not be sufficient or there may be class imbalance in the data set. As such, we have also reduced the rejection threshold to 70\%. The corresponding rejection statistics along with the confusion matrix are shown in Table \ref{tab:tab_incl_rej} and Fig. \ref{fig:confusion_incl_full} respectively. 

\begin{table}
 \centering
 \caption{Precision, recall and $F$ measures for 9-class classification}
 \label{tab:tab_incl_PR}
 \begin{tabular}{>{\bfseries}lcccc }
 \hline
   &  \textbf{Precision (\%)}  &  \textbf{Recall (\%)} & \textbf{$F_{1}$-Score (\%)} & \textbf{Total} \\
 \hline
 $\mathrm{(0^o,15^o)}$ & 1.00 & 1.00 & 1.00 & 509\\
 
 $\mathrm{(0^o,45^o)}$ & 0.91 & 0.95 & 0.93 & 435\\
 
 $\mathrm{(0^o,75^o)}$  & 0.96   & 0.98  & 0.97 &  476\\
 
 $\mathrm{(45^o,15^o)}$ & 0.93 & 0.97 & 0.95 & 430\\
 
 $\mathrm{(45^o,45^o)}$ & 0.98 & 0.99 & 0.99 & 431\\
 
 $\mathrm{(45^o,75^o)}$  & 0.97   & 0.95 & 0.96 &  400\\
 
 $\mathrm{(90^o,15^o)}$ & 0.96 & 0.94 & 0.95 & 504\\
 
 $\mathrm{(90^o,45^o)}$ & 0.96 & 0.91 & 0.94 & 469\\
 
 $\mathrm{(90^o,75^o)}$  & 1.00   & 0.99 & 0.99  &  542\\
 \hline
 Overall &  0.97 & 0.97 & 0.97 & 4196\\
 \hline
 \end{tabular}
\end{table}

\begin{table}
 \centering
 \caption{Number of rejected images from each class in 9-class classification. Rejection threshold is 70\%}
 \label{tab:tab_incl_rej}
 \begin{tabular}{>{\bfseries}lc}
 \hline
  Class &  \textbf{Rejection count}  \\
 \hline
 $\mathrm{(0^o,15^o)}$ & 1\\
 
 $\mathrm{(0^o,45^o)}$ & 75 \\
 
 $\mathrm{(0^o,75^o)}$  & 34  \\
 
 $\mathrm{(45^o,15^o)}$ & 24 \\
 
 $\mathrm{(45^o,45^o)}$ & 23\\
 
 $\mathrm{(45^o,75^o)}$  & 54  \\
 
 $\mathrm{(90^o,15^o)}$ & 51 \\
 
 $\mathrm{(90^o,45^o)}$ & 86\\
 
 $\mathrm{(90^o,75^o)}$  & 13  \\
 \hline
 \end{tabular}
\end{table}

\begin{figure}
    \centering
    \includegraphics[width = \columnwidth]{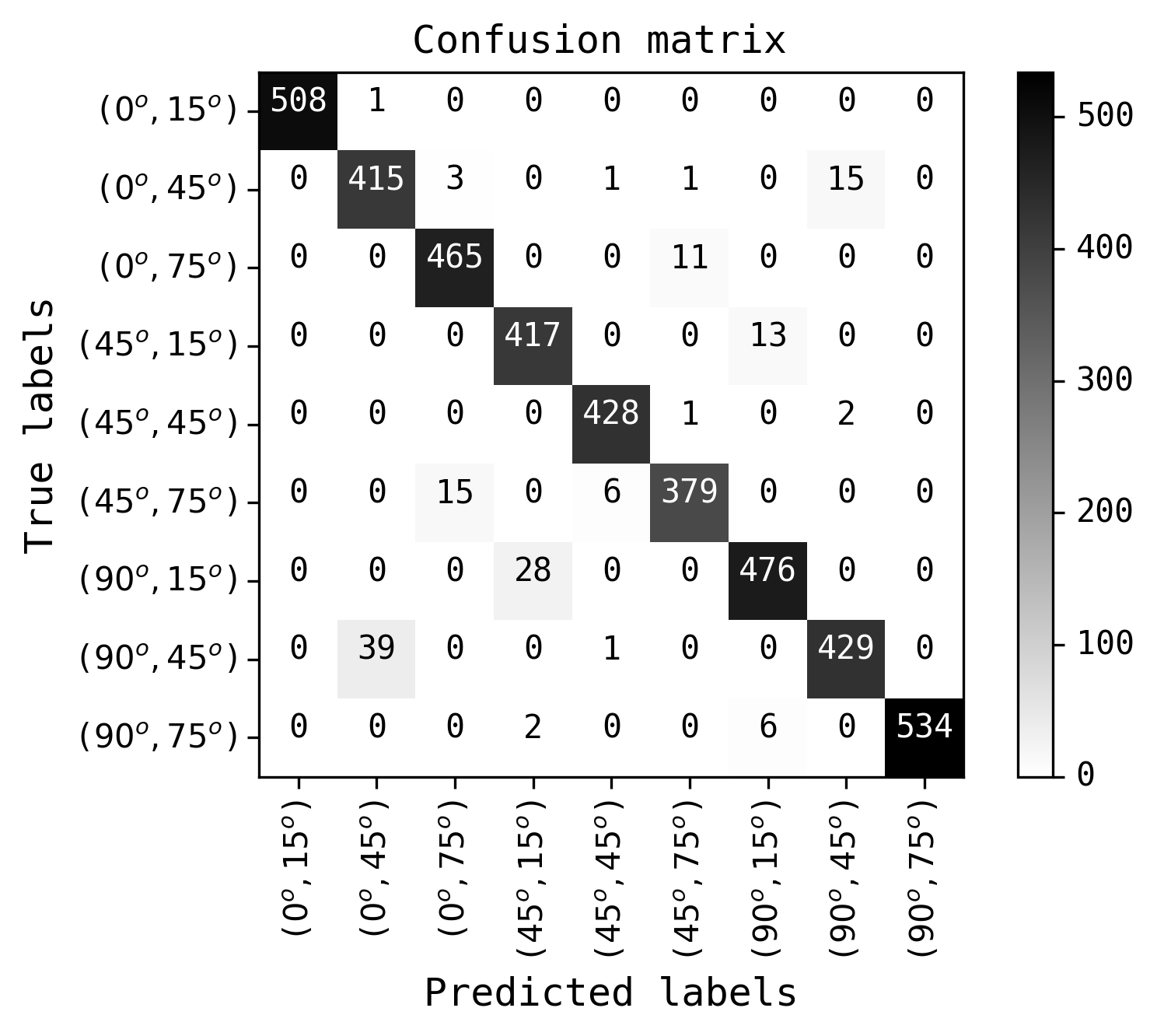}
    \caption{Confusion matrix for 9-class classification.}
    \label{fig:confusion_incl_full}
\end{figure}

We have analyzed several factors that may cause false predictions which generate non-zero off-diagonal values in the confusion matrices. The major factors that can affect the morphological signatures at the close encounter epoch, are relative inclination of the orbital plane of merging pairs, type of orbit (including the total energy of pair and the relative peri-centric distances) and the spin type (prograde or retrograde) of galaxies. Model performance is affected if there are anomalies in the data on spin and orbit combination for  merging pairs.

\subsection*{Case 4: Regression}
\label{sec:sec 5.34}
In GalMer, as continuous $\theta$ values are available for each discrete value of $i$, we have  used CNN based regression architecture, discussed in Section \ref{dynparest}, to estimate them. We have used Adam optimization algorithm \citep{deeplearnebook} to minimize the loss function, mean squared error (MSE), with learning and decay rates of 0.001 and 0.01 respectively. We have collected 181 image samples for each of the integral $\theta$ values in the range, $[-90, 90]$. For our experimentation, we have chosen image samples corresponding to $gS_a - gS_a$ interactions (at their pericentric approach) with $45^\circ$ relative inclination angle ($i = 45^\circ$). However, experiments on other types of interactions (like $gS_a - gS_b$, $gS_b - gS_b$ etc.) and $i$ values (like $0^\circ, 75^\circ \text{ and } 90^\circ$) can also be performed. 

We have used 70\% of 181 samples for training and the remaining 30\% for testing. After each epoch, we have computed MSE on both training and testing data. 

We found that initially, MSE values are high and as the training process continues, they are gradually reduced. After some epochs, training MSE value reduces while testing MSE value increases, due to overfitting. At this stage, we stop the  training process. That means, testing data has been used for validation to avoid overfitting (less error on training data but cannot perform well on unseen testing data). But to achieve good generalization capability, network should exhibit less testing error though training error may be more \citep{duda2012pattern}.  
Finally, an MSE of 0.12 (which is close to 0), has been obtained on testing data. That means, network has achieved a good generalization performance, though the final training MSE value is 0.16.

We note here that both $i$ and $\theta$ are continuous variables, and each of $i$ and $\theta$ can assume any value between $0^{\circ}$ and $90^{\circ}$ for real images. Therefore, ideally, we should have done a full-fledged regression problem, and our CNN should have determined both $i$ and $\theta$ for any given image of an interacting galaxy pair. However, GaLMer, which, to our knowledge, is the only galaxy-merger simulation project which has done a systematic study of the dynamics of an interacting galaxy pair, simulated the same only for a subset of all possible combination of the dynamical and observational parameter values. For example, for the case of 1:1 mass ratio, it has simulated interacting galaxy pairs only for a very few values of $i = 0^{\circ}, 45^{\circ}, 75^{\circ}$ and $90^{\circ}$. Similarly, for the cases of 1:2 and the 1:10 mass ratio, it simulated for 
$i = 33^{\circ}$ only. However, for each of these cases, the GalMer web interface gives the projected image for almost any value of $\theta$ within the allowed range. Therefore, we could not develop a full-fledged regression model to determine both $i$ and $\theta$, but only a partial regression model for the determination of $\theta$ alone for a given value of $i$. The models involving $i$ had to be classification models.

\subsection*{Results on real data}
\label{sec:sec 5.4}
As mentioned earlier, we have trained our CNN model using simulated data from GalMer. We have also applied the trained model for  2-classes, on real data from Sloan Digital Sky Survey (SDSS) Data Release 15 (DR15). DR15 contains all data released upto July 2017 which is the third release of fourth phase of SDSS-IV and it also contains all data from the earlier releases. SDSS Casjobs \footnote{\href{http://skyserver.sdss.org/CasJobs/}{http://skyserver.sdss.org/CasJobs/}} can be used to fetch image data based on right ascension ($RA$) and declination ($DEC$) parameters, using SQL queries from different releases.

SDSS data may have different characteristics as compared to GalMer data. For example, orbital planes of galaxies in the obtained images from SDSS is not known. As such, SDSS images may represent different phenomena or characteristics from GalMer images though $i$ and $\theta$ values are same. But we have still tested our model on only few images just to verify the generalization capability of the network.

To generate the data, we have selectively labelled the images based on user voting such that the manual labelling for at least 80 percent of the users should be consistent. The selected data has then been visually inspected for quality and application feasibility. We have used HyperLEDA \footnote{\href{http://leda.univ-lyon1.fr/}{http://leda.univ-lyon1.fr/}}, a database for galaxies and cosmology, to fetch the individual inclination and position angle for each galaxy in the interacting pair. As discussed in HyperLEDA, position angle is the angle subtended by long axis of galaxy with respect to celestial North and it is measured in terms of 0 to 180 degrees from North to East. Relative inclination angles ($i$) have been computed using Eq~\ref{eq:rel_incl}, based on the individual inclinations ($i_1$ and $i_2$) and position angles (say, $PA_1$ and $PA_2$) of the two galaxies in a pair.  We have used these angles to assign class labels to images. We have shown the above parameters along with SDSS object ID (Obj\_Id) for a few galaxy samples with $0^\circ$ and $45^\circ$ inclination angles in Appendix~\ref{append2} (Tables~\ref{tab:real_0} and \ref{tab:real_45}).  It can be observed from these tables that we have also considered relative inclination values in an interval of $\pm 15^\circ$ around the exact angles ($0^\circ$ or $45^\circ$), as it may not be possible to get more samples only with the exact ones.
\begin{equation}
\centering
\mathrm{
\begin{split}
\label{eq:rel_incl}
\cos (i) & = \sin(i_1)\sin(i_2)\cos(PA_1)\cos(PA_2)  + \\
       & \sin(i_1)\sin(i_2)\sin(PA_1)\sin(PA_2) + \cos(i_1)\cos(i_2)
\end{split}
}
\end{equation}

As SDSS images are collected from real observations, noise may be present in them. So, we have applied median filter on the images, to eliminate noise and later scaled them to a size of $340 \times 340$ pixels. We have tested the final obtained data on our earlier trained binary classification model and the corresponding precision, recall and $F_{1}$ scores are shown in Table~\ref{tab:tab_0_45_real}. From Table~\ref{tab:tab_0_45_real_rej}, it can be seen that 3 image samples are rejected out of 30 images. From the confusion matrix shown in Fig~\ref{fig:con_0_45_real}, it can be observed that 21 image samples are correctly predicted with a confidence level of 80 percent (rejection threshold) while 6 samples are incorrectly predicted. 

We note here that our CNN model is trained on GalMer simulated images of interacting galaxies of mass ratio 1:1 with $i = 0^{\circ}$, $45^{\circ}$, $75^{\circ}$ and $90^{\circ}$. However, in the real scenario, the mass ratio is not restricted to 1:1 and may attain a range of values, even as large as 1:10 or higher. This may raise a concern as to the success of our model in studying real images wherein the mass ratio may be very different from 1:1. However, we have now tested our model on real data to verify the generalization capability of the network and found it to be reasonably successful. Interestingly, the stellar mass ratios of these real interacting galaxies, as deduced from the values of their absolute magnitudes of each galaxy in the interacting pair in different bands (Appendix~\ref{append2}, Tables~\ref{tab:real_0} and \ref{tab:real_45}) spanned a range of values. For example, for the $i=0^{\circ}$ case, our calculations show that the mass ratio mostly lies between 1:1 and 1:8 with a median value of $\sim$ 1:4. For the $i=45^{\circ}$ case, we found that the mass ratio lies between 1:1 to 1:12 with a median value of $\sim$ 1:8. The dynamical mass ratio is expected to lie roughly in the same range as well. This already indicates the success of our model on images corresponding to different mass ratios although it was trained on images with mass ratio 1:1 only.

\begin{table}
 \centering
 \caption{Precision, recall and $F$ measures for 2-class ($0^\circ$ and $45^\circ$) classification on real data.}
 \label{tab:tab_0_45_real}
 \begin{tabular}{ >{\bfseries}lcccc }
 \hline
     &  \textbf{Precision (\%)}  &  \textbf{Recall (\%)} & \textbf{$F_{1}$-Score (\%)} & \textbf{Total} \\
 \hline
 $\mathrm{0^o}$ & 0.82 & 0.69 & 0.75 & 13\\
 
 $\mathrm{45^o}$ & 0.75 & 0.86 & 0.80 &  14\\
 \hline
 Overall &  0.78 & 0.78 & 0.78 & 27\\
 \hline
 \end{tabular}
\end{table}

\begin{table}
 \centering
 \caption{Number of rejected images from each class in 2-class real data classification. Rejection threshold is 80\%}
 \label{tab:tab_0_45_real_rej}
 \begin{tabular}{>{\bfseries}lc }
 \hline
 Class  &  \textbf{Rejection count}  \\
 \hline
 $\mathrm{0^o}$ & 2 \\
 
 $\mathrm{45^o}$  & 1 \\
 \hline
 \end{tabular}
\end{table}

\begin{figure}
 \includegraphics[width=\columnwidth]{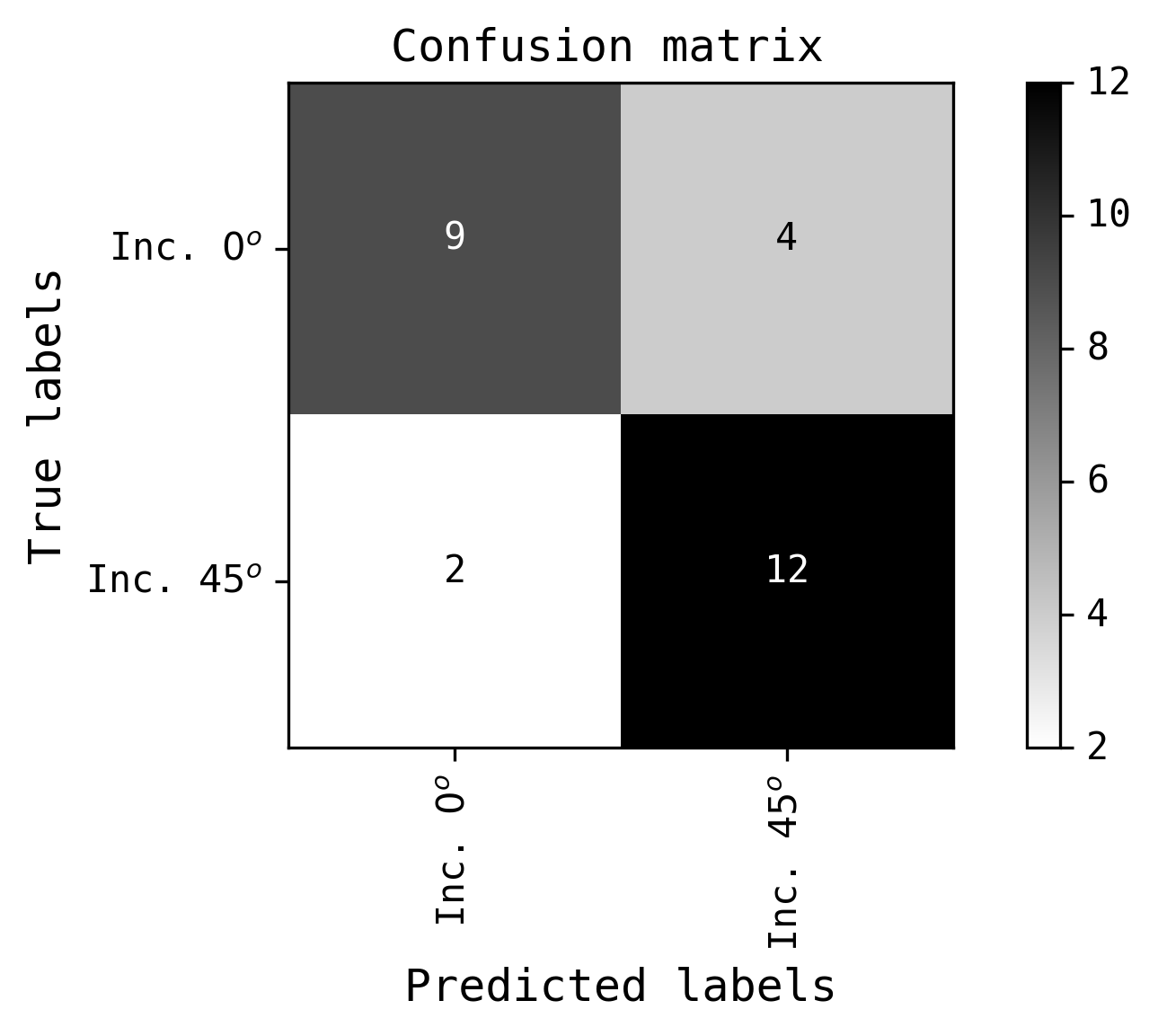}
 \caption{Confusion matrix for 2-class classification on real data.}
 \label{fig:con_0_45_real}
\end{figure}

\section{Conclusions}\label{sec:sec_6}

We have demonstrated the application of Deep Convolutional Neural Networks (DCNN) in determining the fundamental parameters governing dynamical modelling of interacting galaxy pairs:
the relative inclination $i$ between their discs and the viewing angle $\theta$ which is the angle between the line of sight and the normal to the orbital plane. We have used images from GALMER N-body + SPH simulation of 1:1 merger of interacting galaxy pairs at their pericentric approach for training our DCNN models. As GalMer provides discrete values of $i$ and continuous values for $\theta$, we have respectively, used classification and regression to determine them. The training sample represents well galaxy interactions between different morphological types ($gS_a - gS_a$, $gS_a - gS_b$ and $gS_b - gS_b$), also spanning a large range of parameters characterizing the dynamical models as given by the different orbit and spin types. 

Our DCNN model for a (i) 2-class ($i = 0^{\circ}$, $45^{\circ}$ ) (ii) 3-class ($i = 0^{\circ},45^{\circ}, 90^{\circ}$) and (iii) 9-class classification ($(i,\theta) \sim (0^{\circ},15^{\circ})$ ,$(0^{\circ},45^{\circ})$, $(0^{\circ},90^{\circ})$, $(45^{\circ},15^{\circ})$, $(45^{\circ}, 45^{\circ})$, $(45^{\circ}, 90^{\circ})$, $(90^{\circ}, 15^{\circ})$, $(90^{\circ}, 45^{\circ})$, $(90^{\circ},90^{\circ}))$ classification obtained $F_1$ scores of 99\%, 98\% and 97\% respectively. To determine $\theta$, we applied regression model for $i=45^\circ$ from $gS_a - gS_a$ interactions and obtained an MSE of 0.12 (that is close to zero). We also tested our 2-class model on real data from SDSS DR15 and achieved an $F_1$ score of 78\%. Our DCNN models thus can be employed to determine the parameters specifying the initial conditions of dynamical models of galaxy interactions, which is otherwise obtained by an iterative method.

\section*{Data Availability} The raw datasets for training the CNN are available in the GalMer database at \href{http://www.projet-horizon.fr}{http://www.projet-horizon.fr}. For comparison with real images, SDSS Casjobs {\href{http://skyserver.sdss.org/CasJobs/}{http://skyserver.sdss.org/CasJobs/}} can be used to fetch image data based on right ascension ($RA$) and declination ($DEC$) parameters, using SQL queries from different releases.
We have used HyperLEDA {\href{http://leda.univ-lyon1.fr/}{http://leda.univ-lyon1.fr/}}, a database for galaxies and cosmology, to fetch the individual inclination and position angle for each galaxy in the interacting pair.

\section*{Acknowledgements}
We would like to thank Dr. Arun K. Aniyan and Prof. Francoise Combes for useful suggestions and discussion. We would also like to acknowledge DST-INSPIRE Faculty Fellowship (IFA14/PH-101) for supporting this research. We would also like to thank the anonymous referee for his/her useful comments which have improved the quality of the paper.

\addcontentsline{toc}{section}{Acknowledgements}

\bibliographystyle{mnras}
\bibliography{classification}
%%%%%%%%%%%%%%%%%%%%%%%%%%%%%%%%%%%%%%%%%%%%%%%%%%

%%%%%%%%%%%%%%%%% APPENDICES %%%%%%%%%%%%%%%%%%%%%

\appendix

\newpage
\section{Montages}
\label{append1}
For 2-class classification, we have shown a montage of some sample testing images on which, actual and predicted class labels are overlaid, in Figure \ref{fig:mon2}. Figure \ref{fig:mon_0_45_90} shows montage of some sample testing images that include overlaid actual and predicted class labels, for 3-class classification. For 9-class classification, montage of some sample testing images is shown in Figure \ref{fig:mon_incl_full}. Montage of a few sample real images is shown in Fig~\ref{fig:mon_0_45_real}.We note, each of the panels in the above figures is partially grey for reasons of partial contrast enhancement as discussed in Section~\ref{sec:data_preprocessing}. 
\newpage

\begin{figure}
 \includegraphics[width=\columnwidth]{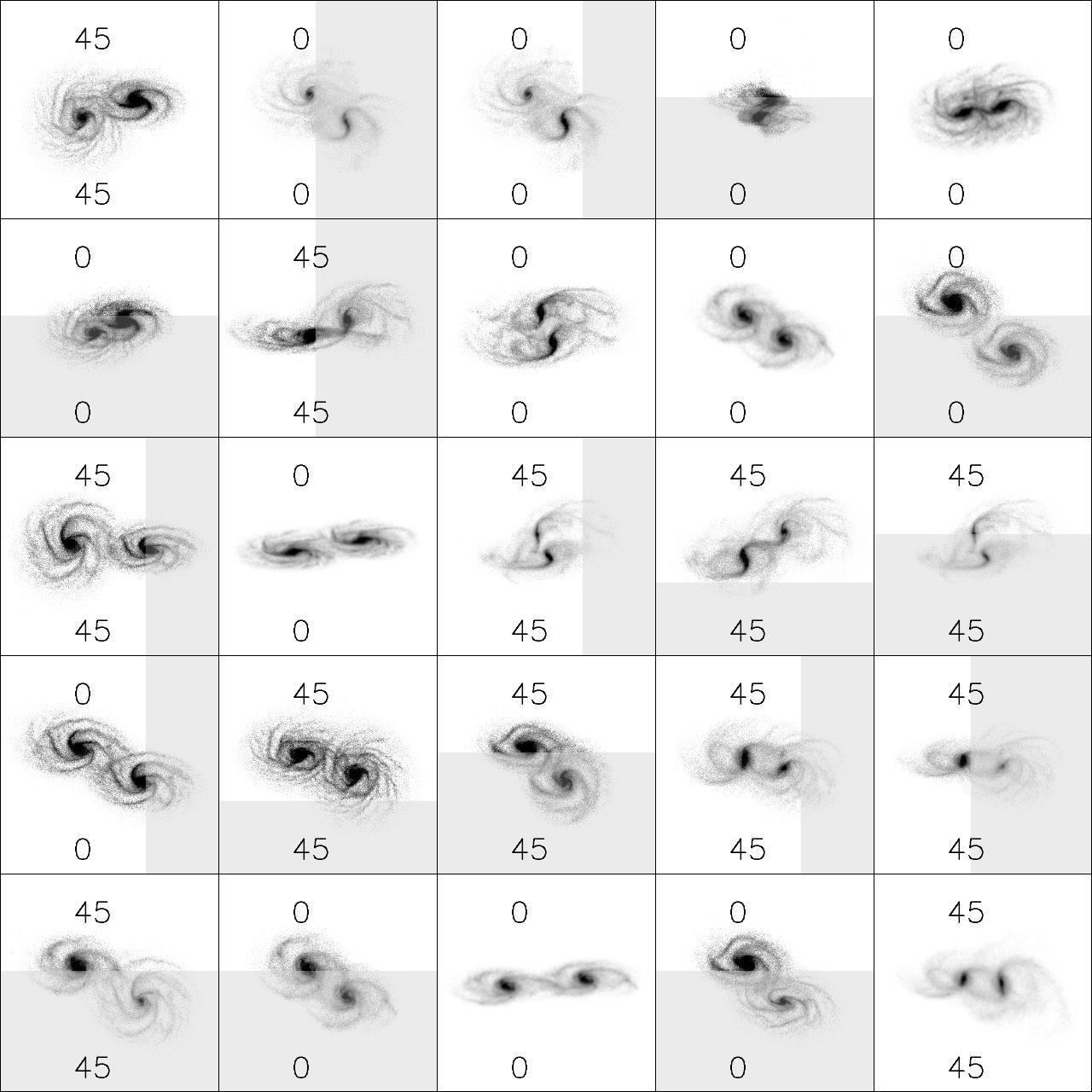}
 \caption{Montage of the training images for 2-class classification. Top and bottom values on each image represent the actual and the predicted class labels respectively.}
 \label{fig:mon2}
\end{figure}

\begin{figure}
 \includegraphics[width=\columnwidth]{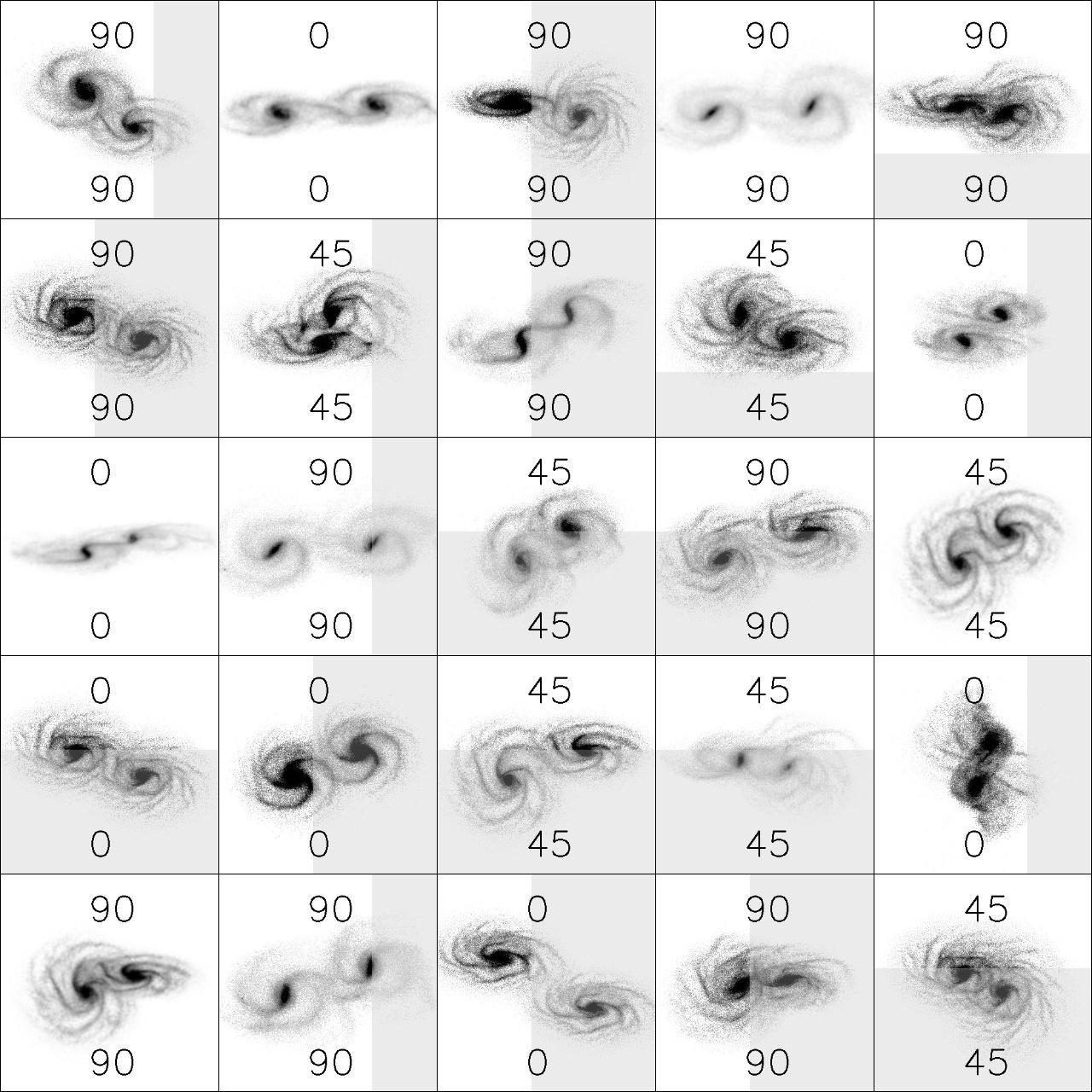}
 \caption{Montage of the training images for 3-class classification. Top and bottom values on each image represent the actual and the predicted class labels respectively.}
 \label{fig:mon_0_45_90}
\end{figure}

\begin{figure}
    \centering
    \includegraphics[width = \columnwidth]{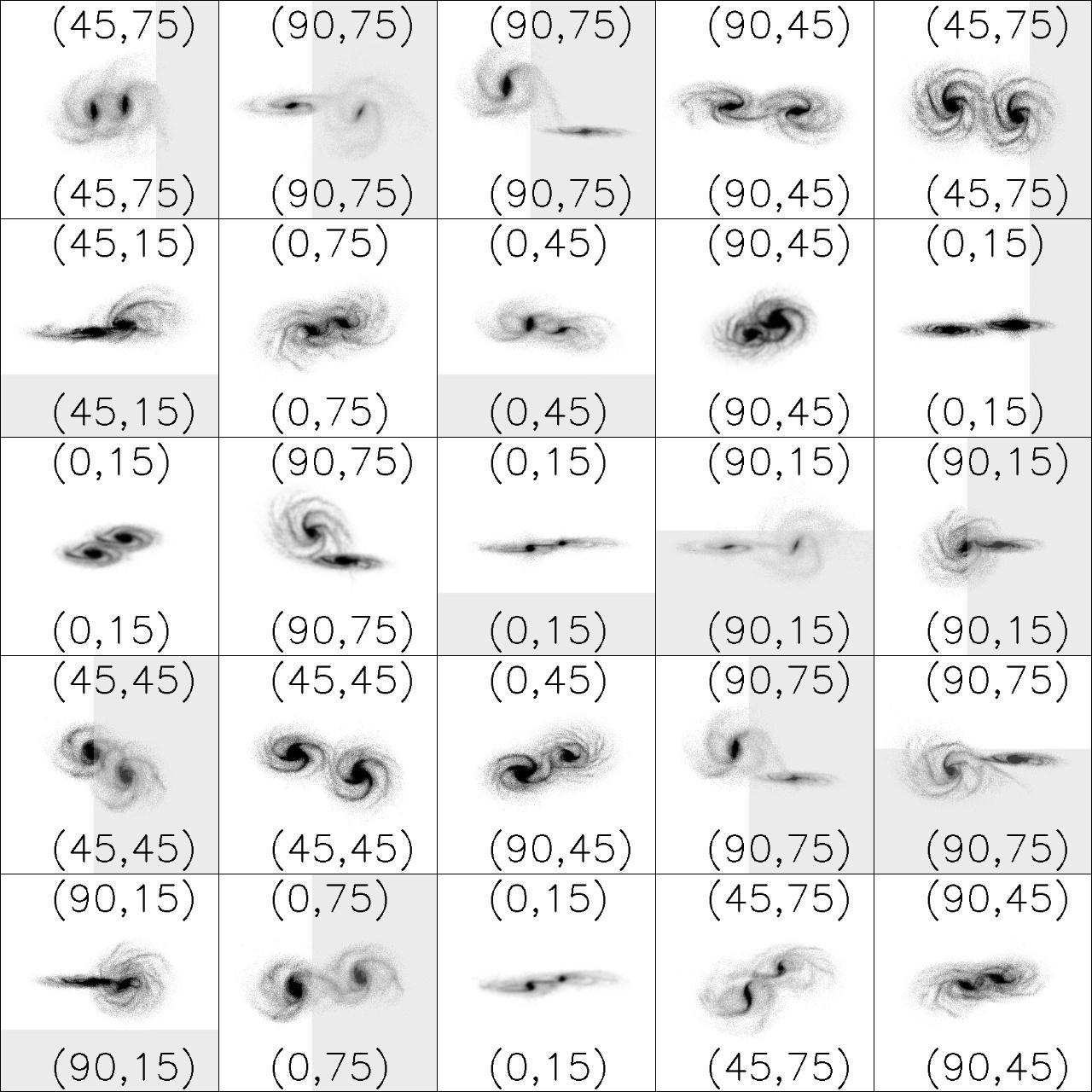}
    \caption{Montage of the training images for 9 class classification. Actual and predicted classes are shown in the top and bottom regions of the image, respectively.}
    \label{fig:mon_incl_full}
\end{figure}

\begin{figure}
 \includegraphics[width=\columnwidth]{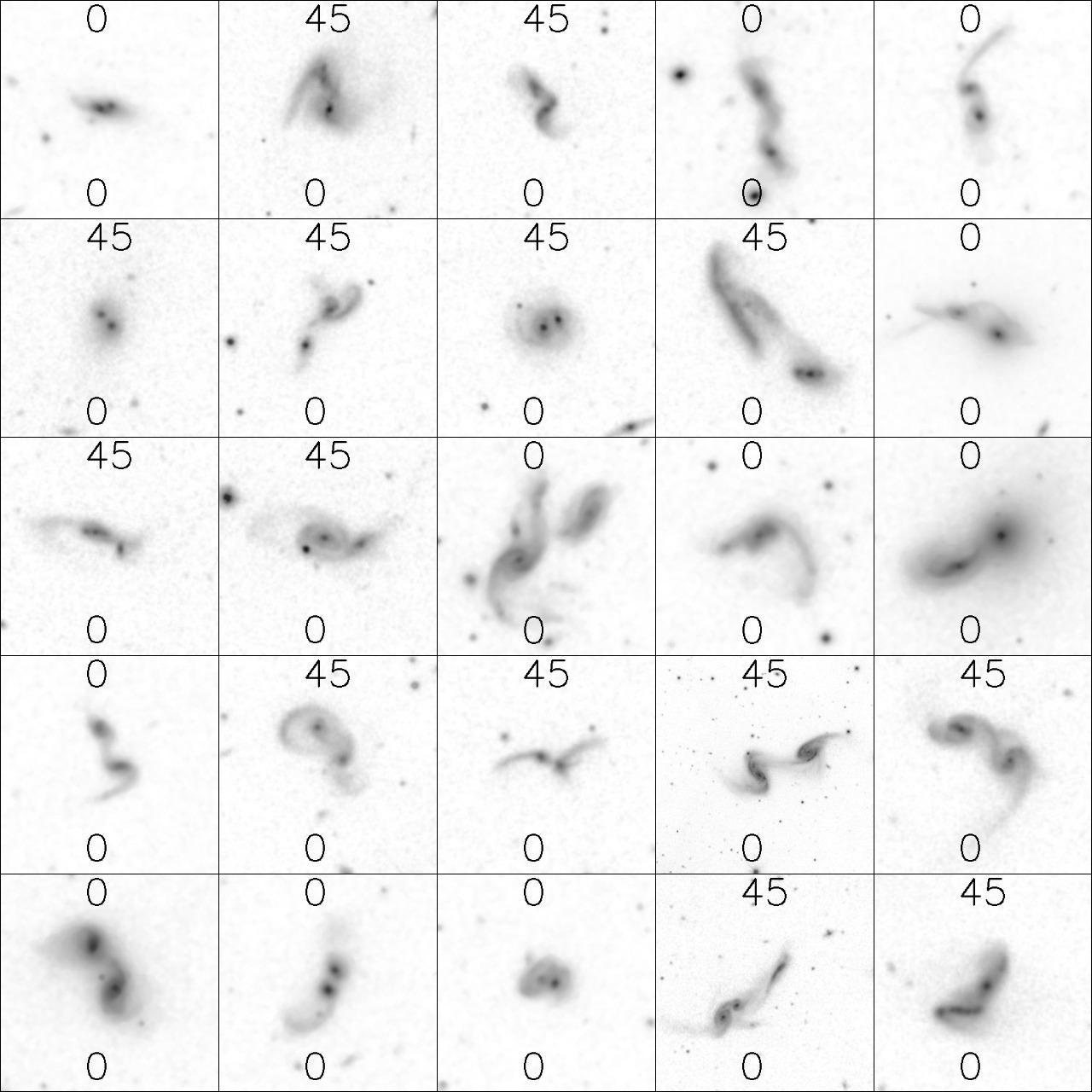}
 \caption{Montage of real images for 0 and 45 degrees interactions. The top and bottom values on each image represent the actual and the predicted class labels.}
 \label{fig:mon_0_45_real}
\end{figure}

\newpage
\section{Galaxy data for different relative inclination angles}
\label{append2}
Different attributes of galaxy data for $0^\circ$ and $45^\circ$ relative inclination angles are shown in Tables~\ref{tab:real_0} and \ref{tab:real_45} respectively. 

\begin{table*}

\begin{minipage}{90mm}
{\small
\hfill{}
\caption{Parameters of a sample of interacting galaxy pairs from observations: Relative inclination angle $\sim$ $0^\circ$ }
\label{tab:real_0}
\hspace*{-3.5cm}
\begin{tabular}{|c|c|c|c|c|c|c|c|c|c|c|c|c|c|}
\hline
Obj\_Id         & RA  \footnote{Right Ascension of the interacting galaxy pair} & DEC \footnote{Declination of the interacting galaxy pair}   & \textbf{$i_1$}\footnote{Angle of inclination of galaxy 1}   & \textbf{$i_2$} \footnote{Angle of inclination of galaxy 2}  & \textbf{$PA_1$} \footnote{Position Angle of galaxy 1}  & \textbf{$PA_2$} \footnote{Position Angle of galaxy 2}   & $i$ & U(1) \footnote{Total $U$-band magnitude of galaxy 1} & B(1)\footnote{Total $B$-band magnitude of galaxy 1}  & I(1) \footnote{Total $I$-band magnitude of galaxy 1} & U(2) \footnote{Total $U$-band magnitude of galaxy 2} & B(2) \footnote{Total $B$-band magnitude of galaxy 2} &  I(2) \footnote{Total $I$-band magnitude of galaxy 2}\\ 
\hline 
J15265746+1009004   & 231.74 & 10.15 & 61.3 & 61.4 & 109.7 & 109.1 & 0.54 & 18.66 &	17.1	&15.47&	19.94&	19.38&	18.41      \\ 
J14312747+1324353   & 217.86 & 13.41 & 59   & 57.7 & 10.6  & 11.1  & 1.37 &  19.7&	18.59&	17.19&	20.69&	19.44&	17.81  \\ 
J00515274-0900558   & 12.97  & -9.02 & 65.4 & 67.4 & 143.2 & 145.6 & 2.97 &   17.49 &	16.5 &	14.76 &	18.31 &	17.21 &	15.85  \\ 
J105953.07+432208.2 & 164.97 & 43.37 & 90   & 90   & 107.1 & 110.6 & 3.50  & 17.97 &	16.12 &	14.84 &	17.73 &	16.15 &	14.69    \\ 
J131530.95+620745.0 & 198.87 & 62.13 & 82.8 & 80.5 & 110.8 & 102.6 & 8.43 & 16.11 &	14.7 &	14.4	 &     16.43 &	15.41 &	14.22      \\ 
J01324722-0825268   & 23.20  & -8.42 & 59.1 & 59.6 & 46.8  & 36.1  & 9.22  &  19.27 &	18.26 &	16.81 &	19.07 &	18.33 &	17.15   \\ 
J092229.43+502548.7 & 140.62 & 50.43 & 90   & 90   & 77.2  & 86.9  & 9.70   &  16.56 &	15.19 &	13.4	 &     16.83 &	15.4 &	14.28   \\ 
J083802.75+075326.2 & 129.51 & 7.89  & 90   & 90   & 107.2 & 117.6 & 10.40 &  19.57 &	16.54 &	16.6	 &     17.41 &	16.12 &	14.23   \\ 
J13485859+1458311   & 207.24 & 14.98 & 77.8 & 67.6 & 20.7  & 25.9  & 11.34  &  18.44 &	17.39 &	16.28 &	19.41 &	17.25 &	16.06  \\ 
J08042267+4038554   & 121.09 & 40.65 & 35.7 & 39.7 & 78.6  & 98.4  & 12.70  &  19.74	 &17.96 &	16.39 &	19.4 &	18	&16.43   \\ 
J104949.74+325903.1 & 162.46 & 32.98 & 66.4 & 57.8 & 29.4  & 40.5  & 13.03  &  13.78	 &13.4 &12.77	 &12.15 &	12.4	&12.06   \\ 
J15013746+2530564   & 225.41 & 25.52 & 52.5 & 56.8 & 139.2 & 154.4 & 13.10  &  19 &	17.3	 &15.75 &	18.7	 &16.89 &	15.35   \\ 
J15063018+6035402   & 226.63 & 60.59 & 50.7 & 40.5 & 29.7  & 17.1  & 13.57  &  17.89	 &17.21 &	16.36 &	17.31 &	16.73 &	15.98  \\ 
J130354.72-030631.6 & 195.98 & -3.11 & 90   & 90   & 157.7 & 171.8 & 14.10   &   19.14	 &17.63 &	16.17 &	20.28 &	18.1	 &15.73  \\ 
J134814.33+152538.5 & 207.06 & 15.43 & 45.5 & 42.5 & 41    & 61.9  & 14.78  &  16.86 &	15.39 &	14.73 &	16.38 &	15.58 &	14.7  \\ \hline

\end{tabular}}
\hfill{}
\end{minipage}
\end{table*}

\begin{table*}
\begin{minipage}{90mm}
{\small
\hfill{}
\caption{Parameters of a sample of interacting galaxy pairs from observations: Relative inclination angle $\sim$ $45^\circ$}
\label{tab:real_45}
\hspace*{-3.5cm}
\begin{tabular}{|c|c|c|c|c|c|c|c|c|c|c|c|c|c|}
\hline
Obj\_Id         & RA  \footnote{Right Ascension of the interacting galaxy pair} & DEC \footnote{Declination of the interacting galaxy pair}   & \textbf{$i_1$}\footnote{Angle of inclination of galaxy 1}   & \textbf{$i_2$} \footnote{Angle of inclination of galaxy 2}  & \textbf{$PA_1$} \footnote{Position Angle of galaxy 1}  & \textbf{$PA_2$} \footnote{Position Angle of galaxy 2}   & $i$ & U(1) \footnote{Total $U$-band magnitude of galaxy 1} & B(1)\footnote{Total $B$-band magnitude of galaxy 1}  & I(1) \footnote{Total $I$-band magnitude of galaxy 1} & U(2) \footnote{Total $U$-band magnitude of galaxy 2} & B(2) \footnote{Total $B$-band magnitude of galaxy 2} &  I(2) \footnote{Total $I$-band magnitude of galaxy 2}\\ 
\hline 
J13563793+5314130   & 209.16 & 53.24 & 64.4 & 90   & 15.8  & 41.9  & 35.92  & 18.42 &	17.62 &	16.28 &	19.63 &	18.49 &	16.96    \\ 
J120423.84+275616.5 & 181.10 & 27.94 & 59.5 & 45.9 & 179.4 & 136.9 & 35.96 &   20.94 &	19.41 &	17.89 &	19.57 &	18.71 &	17.21   \\ 
J133958.60+005006.7 & 204.99 & 0.84  & 34.2 & 69.9 & 177.9 & 166.4 & 36.73   &15.36	 &13.78 &	13.04 &	17.18 &	16.05 &	15.06    \\ 
J125322.35-002542.7 & 193.34 & -0.43 & 46   & 72.7 & 55    & 24.6  & 36.97 &   18.44 &	17.22 &	15.53 &	18.71 &	16.43 &	15.52   \\ 
J11483797+5656514   & 177.16 & 56.95 & 40.8 & 32.1 & 81.6  & 147.9 & 38.66  &  17.64	 &16.65 &	15.35 &	18.42 &	17.39 &	15.87   \\ 
J10174066+2047438   & 154.42 & 20.80 & 42.2 & 63.8 & 169.8 & 126.4 & 40.09 &   20.73 &	18.94 &	17.23 &	19.14 &	17.9 &	16.55   \\ 
J11263223+4102333   & 171.63 & 41.04 & 41   & 38.5 & 158.2 & 91.5  & 41.22   & 19.28	 &18.19 &	16.66 &	19.72 &	18.91 &	17.22   \\ 
J120439.56+525726.6 & 181.16 & 52.95 & 57.3 & 67.7 & 41.6  & 87.9  & 42.02  &   18.02	 &17.1 &	15.59 &	18.37 &	17.15 &	15.7  \\ 
J13442640-0224196   & 206.11 & -2.41 & 67.9 & 82.4 & 34.2  & 76    & 42.74    &   20.44 &	18.84 &	17.39 &	20.11 &	17.68 &	16.41\\ 
J08281380+1741338   & 127.06 & 17.69 & 37.2 & 28.9 & 117.6 & 31.2  & 44.30  &   18.57	 &17.29 &	15.74 &	19.83 &	18.22 &	16.64  \\ 
J133614.07-010217.3 & 204.05 & -1.03 & 70.9 & 36.1 & 152.6 & 115   & 45.13   &    17.18 &	15.76 &	14.15 &	16.74 &	15.62 &	14.94\\ 
J14165155+5410476   & 214.21 & 54.18 & 53.6 & 90   & 56.9  & 87.3  & 46.03   &    19.34 &	18.78 &	17.58  &	19.97 &	18.8	 &17.3\\ 
J08333938+5152066   & 128.41 & 51.87 & 90   & 38.9 & 2.8   & 3.8   & 51.11    &   20.99	 &18.46 &	16.54 &	21.51 &	18.93 &	16.93\\ 
J08513476+3916087   & 132.89 & 39.27 & 58.6 & 52.3 & 97.7  & 160.5 & 51.15  & 17.21	 &16.46 &	15.47 &	18.41 &	17.26 &	15.82    \\ 
J09543272+5633246   & 148.64 & 56.56 & 90   & 51   & 106.8 & 66.2  & 53.84    &  19.71	 &18.15 &	16.65 &	18.39 &	17.19 &	15.93 
\\ \hline
\end{tabular}}
\hfill{}
\end{minipage}
\end{table*}

%%%%%%%%%%%%%%%%%%%%%%%%%%%%%%%%%%%%%%%%%%%%%%%%%%

% Don't change these lines
\bsp	% typesetting comment
\label{lastpage}
\end{document}